\PassOptionsToPackage{desactivate}{linenoaa}
\documentclass{aa}  
\usepackage{graphicx}
\usepackage{txfonts}
\usepackage{lipsum}
\usepackage{subcaption}
\usepackage{lscape}
\usepackage{placeins}
\usepackage{booktabs}
\usepackage{graphicx}
\usepackage[dvipsnames]{xcolor}
\usepackage{tikz}
\usetikzlibrary{positioning}
\usepackage{float}
\usepackage{stfloats}
\usepackage{txfonts}
\usepackage{natbib}
\bibpunct{(}{)}{;}{a}{}{,}

\usepackage{hyperref}

\begin{document} 
   \title{Estimating the completeness of the QUBRICS Survey with 3501 QSO redshifts from Gaia DR3 spectra}

   \author{Matteo Porru
          \inst{1}\fnmsep\thanks{E-mail: matteo.porru@inaf.it},
          Stefano Cristiani     \inst{1,2,3},
          Francesco Guarneri    \inst{11,1},
          Giorgio Calderone     \inst{1},
          Andrea Grazian        \inst{4},
          Konstantina Boutsia   \inst{5},
          Andrea Trost          \inst{1,3,6},
          Valentina D'Odorico   \inst{1,2,12},
          Guido Cupani          \inst{1,2},
          Catarina M.J. Marques \inst{7,8,9},
          Francesco Chiti Tegli \inst{10},
          Fabio Fontanot        \inst{1},
          }

     \institute{
     INAF--Osservatorio Astronomico di Trieste, Via G.B. Tiepolo, 11, I-34143 Trieste, Italy
     \and
     IFPU--Institute for Fundamental Physics of the Universe, via Beirut 2, I-34151 Trieste, Italy
     \and
     INFN--National Institute for Nuclear Physics, via Valerio 2, I-34127 Trieste, Italy
     \and
     INAF--Osservatorio Astronomico di Padova, Vicolo dell'Osservatorio 5, I-35122 Padova, Italy
     \and
     Cerro Tololo Inter-American Observatory/NSF NOIRLab, Casilla 603, La Serena, Chile
     \and
     Dipartimento di Fisica dell’Università di Trieste, Sezione di Astronomia, Via G.B. Tiepolo, 11, I-34143 Trieste, Italy
     \and
     Centro de Astrofísica da Universidade do Porto, Rua das Estrelas, 4150-762 Porto, Portugal
     \and
     Instituto de Astrofísica e Ciências do Espaço, Universidade do Porto, Rua das Estrelas, 4150-762 Porto, Portugal
     \and
     Faculdade de Ciências, Universidade do Porto, Rua do Campo Alegre, 4150-007 Porto, Portugal
     \and
     Max-Planck-Institut für Astronomie, Königstuhl 17, D-69117 Heidelberg, Germany
     \and
     Hamburger Sternwarte, Universität Hamburg, Gojenbergsweg 112, D-21029 Hamburg, Germany
     \and
     Scuola Normale Superiore Piazza dei Cavalieri, 7, I-56126 Pisa, Italy
     }
   \date{Received ...; accepted ...}

\titlerunning{The completeness of the QUBRICS Survey} 
\authorrunning{Porru et al.}
  \abstract
   {Quasi-Stellar Objects (QSOs) are essential for investigating the structure and evolution of the Universe. Historically, their identification has been concentrated in the northern hemisphere, primarily due to the sky coverage of major astronomical surveys. The QUBRICS (QUasars as BRIght beacons for Cosmology in the Southern hemisphere) survey, started in 2019 to address this asymmetry, has identified more than 1300 new bright ($i<19.5$) high-redshift ($2.5<z<6$) QSOs in the southern sky.} 
   {This study aims to quantify, using an independent QSO sample, the completeness and recall of the QUBRICS QSO selection methods, based on XGB (eXtreme Gradient Boosting) and PRF (Probabilistic Random Forest), since completeness is a fundamental metric for ensuring the statistical robustness of QSO-based cosmological investigations.}
   {A subset ($G<18.25$, $|b|>25$ deg, negligible parallax and proper motion) of Gaia DR3 sources with low-resolution spectra was analyzed, obtaining a sample of 3501 QSOs.
   To determine how many QSOs were correctly identified as candidates, we crossmatched this independent sample with the datasets used for selection: 894 QSOs with $z>2.5$ fell within the XGB dataset footprint, of which 152 were unclassified and thus eligible for completeness testing. Similarly, 675 QSOs with $z>2.5$ were within the PRF dataset footprint, including 69 unclassified objects. 
   }
   {
   The XGB correctly identified as candidates 136 (89\%) of the 152 QSOs with $z>2.5$ present in the XGB dataset as unclassified objects.
   The PRF correctly identified as candidates 46 (66\%) of the 69 QSOs with $z>2.5$ present in the PRF dataset as unclassified objects.
   }
   {These findings confirm the high efficiency of the QUBRICS selection methods (recall $=89\%$) and provide the completeness estimate for spectroscopically confirmed QSOs (82\%), necessary for cosmological studies using QUBRICS data. This work also provides reliable redshifts for 1223 new QSOs (median redshift $z=2.1$ and magnitude $G=17.8$), that will help improve the performance of future selections.}

   \keywords{-- methods: data analysis -- methods: statistical -- surveys -- astronomical databases: miscellaneous -- quasars: general
               }

   \maketitle

\section{Introduction}
\label{sec:introduction}
Surveys of Quasi-Stellar Objects (QSOs) play a key role in advancing our understanding of the structure and evolution of the Universe. By studying the distribution and properties of QSOs, it is possible to gain insights into the nature of dark matter, dark energy, and other foundational aspects of physics, as well as testing current theories of cosmology.
Light from distant and powerful QSOs has proven to be an invaluable tool to probe the Inter-Galactic Medium  \citep[IGM, e.g.][]{Meiksin2009,McQuinn2016}, to investigate the possible variation along cosmic time of the fundamental constants of
nature \citep[e.g.][]{Murphy2022}, to study the development and the sources of the Reionization process \citep[e.g.][]{Bosman2022, Dayal2024, Madau2024}, and even allow a direct measurement of the redshift drift, i.e., the temporal variation of cosmological redshifts as a probe of the Universe expansion \citep[e.g.][]{Liske+08:2008MNRAS.386.1192L}.
These cosmological beacons are expected to become increasingly important in the coming decade, with the upcoming availability of high-resolution spectrographs on 40m-class telescopes (e.g., ANDES/ELT, \citealt{ANDES_2024}) that will enable the detection of the redshift drift \citep{Trost2025, Trost2026}. 
QSO surveys also enable us to probe the earliest stages of galaxy formation, the formation and evolution of supermassive black holes \citep[e.g.][]{Trakhtenbrot2021}, and shed light on the processes that shape the Universe we observe today \citep[e.g.][]{Fan2023}. Catalogs with a high completeness of confirmed bright QSOs at high redshift are therefore essential to achieving these goals.

Surveys covering large portions of the sky (e.g., SDSS, \citealt{LykeSDSS16q:2020ApJS..250....8L}) have proven very efficient in finding relatively bright QSOs at high redshift. However, these projects have later been shown to be incomplete by 30-40\% by smaller-scale surveys \citep[e.g.][]{SDSSIncomplete_Schindler:2019ApJ...871..258S, LF_Boutsia:2021ApJ...912..111B} at high redshift and bright magnitudes ($z\ge2.8$, $i\le18$), suggesting that a significant fraction of bright QSOs in the distant Universe is still undetected. Determining the completeness of QSO surveys is a critical step in achieving the aforementioned goals (e.g., computing the luminosity function) and ensuring that our understanding of the Universe is not biased. 

In this paper, we test and discuss the completeness of the QUBRICS (QUasars as BRIght beacons for Cosmology in the Southern hemisphere) survey \citep{Calderone19:2019ApJ...887..268C}.
QUBRICS is a spectroscopic survey targeting bright ($i<19.5$) high-redshift ($z>2,5$) QSOs, with the primary goal of addressing the scarcity of bright QSOs in the southern sky caused by the historical paucity of all-sky surveys in the South. 
The survey utilizes photometry from the datasets:
\begin{itemize}
    \item Gaia DR3 \citep{GaiaDR3:2023A&A...674A...1G}
    \item Panoramic Survey Telescope and Rapid Response
System 1 DR2 \citep[PanSTARRS1, ][]{PanSTARRS:Chambers_2016}
    \item SkyMapper Southern Survey DR4 \citep[SMSS, ][]{SkyMapper4:2024PASA...41...61O}
    \item Dark Energy Survey DR2 \citep[DES, ][]{DESDR2:2021ApJS..255...20A}
    \item AllWISE\footnotemark{} \citep{AllWISE_2014yCat.2328....0C}
    \item CatWISE2020\footnotemark[\value{footnote}] \citep{CatWISE:2021ApJS..253....8M}
\end{itemize}

\footnotetext{The AllWISE and CatWISE catalogs combine data from the Wide-field Infrared Survey
Explorer \cite{WISE:2010AJ....140.1868W} and Near-Earth Object Wide-Field Infrared Survey Explorer \cite{NEOWISE:2014ApJ...792...30M} missions.}

Magnitudes for PanSTARRS, SkyMapper and DES are in the AB magnitude system, while Gaia, AllWISE and CatWISE magnitudes are in the Vega system.

The QUBRICS database also integrates spectroscopic classifications and redshifts from primary catalogs (SDSS DR16Q, \citealt{LykeSDSS16q:2020ApJS..250....8L}, 2dFGRS, \citealt{2df:2001MNRAS.328.1039C}, DESI, \citealt{DESI-EDR1:2024AJ....168...58D}) and several publications \citep[e.g.][]{Veron10:2010A&A...518A..10V, Bosman:2020zndo...3634964B, Wolf20QSO:2020MNRAS.491.1970W, Onken2022:2022MNRAS.511..572O, Yang2023:2023ApJS..269...27Y} as well as our own follow-up campaigns. 

By using machine learning methods trained on these datasets to perform photometric candidate selection, QUBRICS observation campaigns have so far produced more than a thousand new spectroscopically confirmed bright QSOs at $z>2.5$. 
Various methods for the selection of QSOs have been used: 
in \citet{Calderone19:2019ApJ...887..268C} candidates have been selected using Canonical Correlation Analysis \citep[CCA, ][]{ref:CCA}, while in \citet{Guarneri:2021MNRAS.506.2471G} the Probabilistic Random Forest \citep[PRF, ][]{ReisPRF:2019AJ....157...16R} algorithm has been adopted, with modifications introduced to properly treat upper limits and missing data.
In \citet{Guarneri2022} the PRF selection has been further improved, adding synthetic data to the training sets. 
\citet{Calderone2024} have developed a method that takes advantage of the extreme gradient boosting technique \citep[XGB,][]{XGBoost2016} to significantly improve the recall of the selection algorithms (from 54\% to 86\%) even in the presence of severely imbalanced datasets, with the aim of extending the redshift range of the QUBRICS survey up to $z \sim 5$.

In addition to the continuous improvement of the photometric selection methods, a consistent effort has been dedicated to follow-up spectroscopy \citep{Boutsia2020,Cristiani2023}. Observation runs under QUBRICS have provided reliable spectroscopic classification and redshifts for 1793 previously unidentified objects, of which 1659 (92.5\%) are confirmed QSOs, with 1342 (74.8\%) having a spectroscopic redshift above 2.5. The magnitude range covered is $16<i<19.5$ and $-29.5<M_{1450}<-26$, while the redshift range covered is $2.5<z<5.8$.
These results confirm the effectiveness of the selection procedures, as well as providing well-defined subsamples with statistically high completeness that allowed us to address the topics of the QSO luminosity function (LF) and cosmic re-ionization(s)
\citep{LF_Boutsia:2021ApJ...912..111B, Grazian2022, Fontanot2023}.
Rare objects, such as extreme broad absorption line QSOs (BALs), discovered in the course of QUBRICS, have also been a topic of study in \citet{ref:cupaniFeLoBALs}.

In this paper, we aim to estimate the completeness and recall of the two main candidate selection methods, based on the XGB and PRF algorithms.

The paper is structured as follows:
Sect. \ref{sec:define_qso_sample} describes how the independent QSO sample was created, starting from the analysis of the Gaia DR3 spectra;
Sect. \ref{sec:validate_qso_sample} describes how the QSO sample was validated by comparing the redshifts of known QSOs; 
Sect. \ref{sec:XGB} describes how the QSO sample was used to measure the completeness and recall of the XGB selection; 
Sect. \ref{sec:PRF} describes how the QSO sample was used to measure the completeness and recall of the PRF selection; 
Sect. \ref{sec:Reliability} discusses the reliability of the recall estimates; 
Sect. \ref{sec:conclusions} analyzes the obtained estimates for the completeness and discusses the results.

\section{Defining an independent QSO sample to test the completeness of QUBRICS}
\label{sec:define_qso_sample}

In this paper, we aim to measure the completeness of two selection methods of QSO candidates that are central to the QUBRICS survey: the PRF-based selection, described in \citet[hereafter Paper I]{Guarneri2022}, and the reverse-selection method, based on the XGB algorithm, described in \citet[hereafter Paper II]{Calderone2024}.
While estimates have been produced in the context of the respective candidate selections, it is important to carry out independent assessments to ensure a reliable determination of the completeness, which is needed for the achievement of many scientific goals of QUBRICS, as well as for understanding the causes of the incompleteness.

We adopt the following definitions:
\begin{itemize}
    \item Dataset completeness: the number of true QSOs in the dataset (with or without a classification) divided by the number of true QSOs in the sky;
    \item Spectroscopic completeness: the number of spectroscopically confirmed QSOs in the dataset divided by the number of true QSOs in the sky;
    \item Selection recall: the number of predicted QSOs in the dataset that are actually QSOs divided by the number of true QSOs that have no classification in the dataset;
    \item Selection completeness: the number of predicted QSOs in the dataset that are actually QSOs divided by the number of of true QSOs in the sky that have no classification.
\end{itemize}
For example, consider a region of the sky with 100 QSOs in a given magnitude range.
If we have a dataset where 90 of them have been detected as sources (i.e., we know their coordinates and photometric magnitudes), then the dataset completeness is 90\%.
If our dataset has spectroscopic classification for 30 of them, then its spectroscopic completeness is 30\%. If we apply a selection algorithm on the unclassified fraction of the dataset and identify 40 of the 60 unclassified QSOs in the dataset, the recall is 66\%, and the selection completeness is 60\% (i.e., the recall times the dataset completeness). 
Thus, completeness is a measure of how comprehensive a dataset is, while recall is a measure of how good a selection algorithm is at finding all the objects of interest.
If all the candidates found by the selection algorithm are observed, the resulting dataset of new objects will have a spectroscopic completeness equal to the selection completeness.

To measure these quantities, an independent QSO sample has been created, based on the Gaia DR3 main catalog  \citep{GaiaDR3:2023A&A...674A...1G}. The candidates had to satisfy the following criteria:
\begin{enumerate}
    \item having an associated Gaia DR3 low-resolution BP/RP spectrum;
    \item negligible proper motion ($PM/PM_{err}<3$);
    \item negligible parallax ($parallax\_over\_error<3$);
    \item apparent magnitude $14<G<18.25$\footnote{The magnitude range $14<G<18.25$ was chosen pragmatically to keep the visually-inspected sample manageable (resource‑driven choice), while still focusing on the brightest sources with an acceptable SNR.};
    \item galactic latitude $|b|>25 \deg$.
\end{enumerate} 
The Gaia DR3 low-resolution BP/RP spectra cover wavelength range 330-1050 nm with spectral resolution R$\sim$20-100 \citep{CarrascoGaiaSpectra:2021A&A...652A..86C}, sufficient to identify broad emission lines such as $Ly\alpha$ (1216 $\AA $), \ion{Si}{IV} (1397 $\AA $) and \ion{C}{IV} (1549 $\AA $) at $z>2.5$.
The spectra of the resulting 37504 sources have been analyzed with the method described in Sect. 3.2 of \citet{Cristiani2023}: we adopted the Marz software\footnote{\url{https://samreay.github.io/Marz/}} \citep{MARZ_2016}, which uses a cross-correlation algorithm to determine the redshift. Each spectrum is compared to a QSO template designed specifically for QSO identification and containing the emission features characteristic of QSO spectra. The algorithm computes the correlation coefficient between the spectrum and the template across the $0<z<5$ redshift range; the redshift corresponding to the maximum correlation value is then returned as the best-fit redshift estimate. After the automatic processing with the Marz software, visual inspection was performed, in order to assign to each spectrum a subjective quality rating (QOP) and to adjust the redshift estimate when necessary (e.g., when a second cross-correlation peak provides a better agreement between the spectrum and template emission lines).

\begin{table*}[h!]
\centering
\caption{The 12 highest redshift QSOs discovered in this work.}
\begin{tabular}{|r|c|c|c|c|c|c|}
\hline
\textbf{qid} & \textbf{RA} & \textbf{DEC} & \textbf{Gmag} & \textbf{QOP} & \textbf{z\_QU\_G} \\ \hline
42587316 & 15:50:19.82 & +63:22:25.4 & 18.06 & 3 & 3.804 \\ \hline
42579561 & 12:34:01.53 & +17:54:09.2 & 18.16 & 3 & 3.522 \\ \hline
42587168 & 07:32:57.28 & +54:52:11.6 & 17.96 & 4 & 3.384 \\ \hline
42587209 & 07:08:02.48 & +63:15:59.7 & 17.40 & 4 & 3.325 \\ \hline
42587278 & 15:39:23.22 & +35:15:36.2 & 18.09 & 2 & 3.305 \\ \hline
42587128 & 02:35:01.80 & +22:55:48.7 & 17.66 & 3 & 3.249 \\ \hline
42587862 & 00:12:46.01 & -62:51:51.5 & 18.14 & 4 & 3.248 \\ \hline
1016233  & 04:24:14.33 & -22:35:00.3 & 18.00 & 3 & 3.244 \\ \hline
42587124 & 03:28:33.54 & +15:38:44.1 & 17.84 & 3 & 3.221 \\ \hline
42587634 & 13:21:04.76 & -17:04:39.4 & 18.19 & 3 & 3.192 \\ \hline
42587486 & 23:55:29.65 & +09:29:30.0 & 18.19 & 2 & 3.174 \\ \hline
42587467 & 01:36:30.19 & +17:20:23.2 & 18.22 & 3 & 3.172 \\ \hline
\end{tabular}
\tablefoot{The corresponding spectra are shown in Fig. \ref{Fig:qso_sample_spectra}). Coordinates and G magnitudes are from Gaia DR3. A higher QOP indicates a better quality spectrum (QOP=2: acceptable, QOP=3: good, QOP=4: great). The full table with 3501 QSOs is available at the \href{https://cds.unistra.fr/}{CDS}.}
\label{tab:qso_sample}
\end{table*}

The main criteria taken into account were the maximum value of the cross-correlation between spectrum and QSO template, the signal-to-noise ratio (SNR) and the number of visible features. The resulting QOP ranged between 1 (bad: uncertain identification and/or redshift) and 4 (great: certain identification and redshift). To limit subjectivity, QOP assignments were performed following the same criteria used in \cite{Cristiani2023} and are illustrated by representative examples in Appendix \ref{appendix:gaia-spectra-examples}.

Considering the redshift range of interest for QUBRICS ($z>2.5$) and the wavelength range of Gaia spectra ($330-1050$ nm), we focused on spectra that showed $Ly\alpha$ (1216 $\AA $), \ion{Si}{IV} (1397 $\AA $) and \ion{C}{IV} (1549 $\AA$) emission lines clearly identifiable by visual inspection. This process resulted in 3501 objects with Gaia spectra of sufficient quality (QOP$\ge2$) to be identified as QSOs and measure a redshift, that corresponds to around 10\% of all analyzed bright ($G<18.25$) spectra. Some examples of Gaia spectra are shown in Fig. \ref{Fig:qso_sample_spectra}: despite being low-resolution, the spectra show clear emission lines that enable QSO identification.
We prioritized spectral reliability by excluding uncertain objects; as a result, the sample is robust, though not fully complete. Thus, the sample does not represent the full QSO population in Gaia DR3, which is acceptable for evaluating the completeness, but should be kept in mind when interpreting the results.
Since the Gaia QSO sample was selected independently from the XGB and PRF samples, we can reasonably assume that no correlation exists between the respective sources of incompleteness and that the estimate of the completeness of the XGB and PRF selection is not affected by the incompleteness in the independent sample.
Therefore, we do not assess the completeness of the Gaia QSO sample itself, but focus exclusively on evaluating the completeness of the XGB and PRF selections relative to it.

The coordinates, magnitudes and $z_{\rm QU\_G}$ redshifts of these QSOs are listed in Table \ref{tab:qso_sample} (full table available in electronic form). Fig. \ref{Fig:qso_sample_spectra} shows the Gaia spectra for the 12 QSOs with the highest redshift identified in this work.
Of the 3501 QSOs, 1115 have $z>2.5$ (the redshift range of interest for QUBRICS).

\begin{figure}[h!]
    \centering
    \includegraphics[width=\columnwidth]{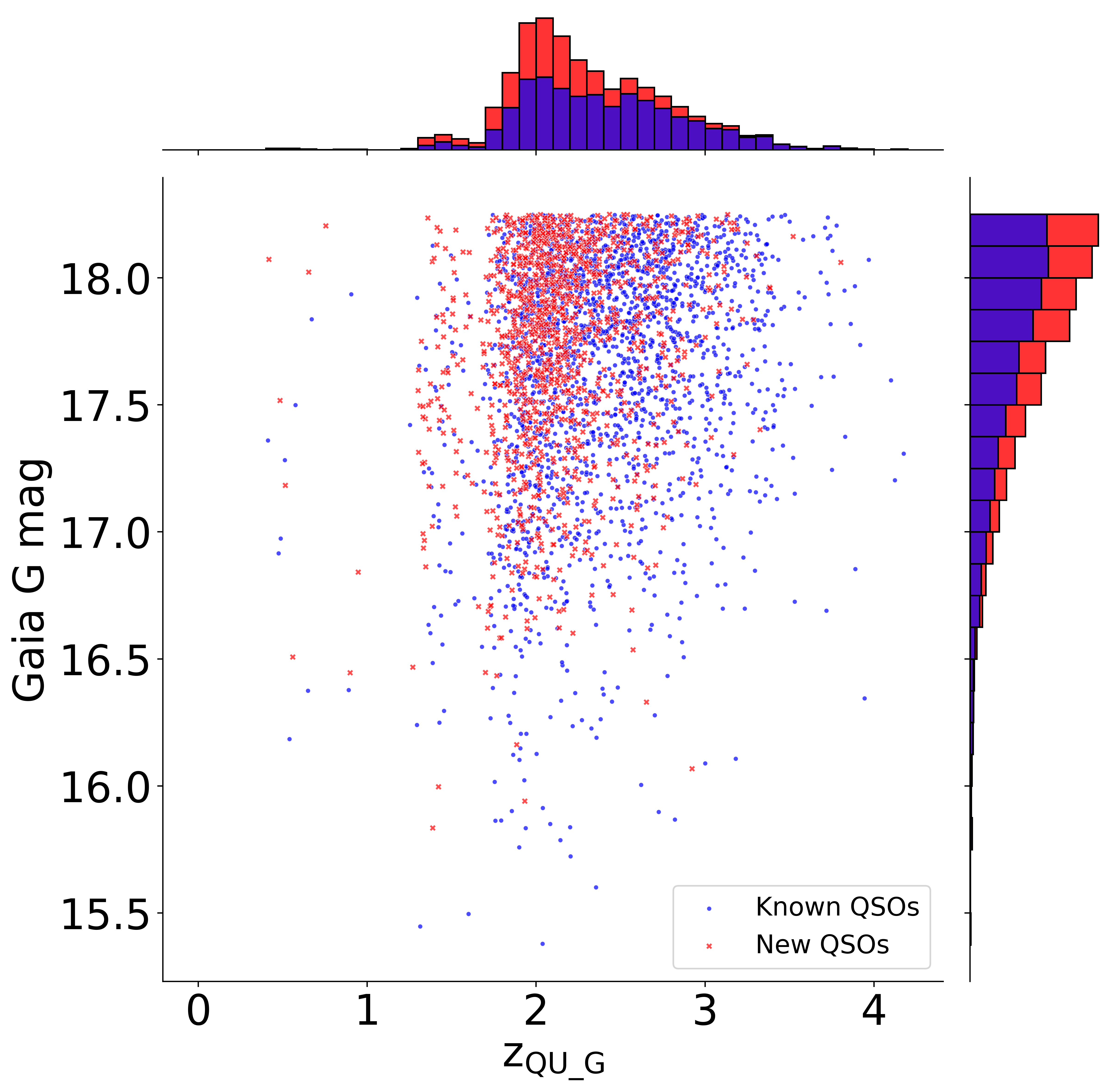}
    \caption{The Gaia G magnitude of the 3501 QSOs of the independent sample plotted vs the $z_{\rm QU\_G}$ redshifts determined on the basis of the Gaia low-resolution spectra. 2278 previously known QSOs are highlighted in blue, while 1223 newly identified QSOs are highlighted in red. }
    \label{Fig:Gmag_vs_z_marz}
\end{figure}

In Fig. \ref{Fig:Gmag_vs_z_marz} the distribution of the Gaia G magnitude vs. the redshift determined on the basis of the Gaia low-resolution spectra is shown. The distribution appears to disfavor QSOs below $z \sim 2$, which is likely a selection effect due to the wavelength range of Gaia DR3 spectra: the $Ly\alpha$ emission line enters that specific spectral range only around $z \sim 1.8$.

\section{Validating the independent QSO sample}
\label{sec:validate_qso_sample}
In order to validate the sample of 3501 QSOs obtained from the Gaia DR3 spectra, we compared it with the objects having a spectroscopic classification in the QUBRICS database. A coordinate crossmatch has been performed, using a maximum radius of 0.5 arcseconds to avoid false matches. 2278/3501 (65\%) of the objects in the sample were found to be classified as QSOs in the QUBRICS database (as of August 2024). 
Considering the redshift range of interest for QUBRICS ($z>2.5$), 910/1115 (82\%) of the objects in the sample with $z_{\rm QU\_G}>2.5$ were already classified as QSOs in the QUBRICS database. The spectroscopic completeness of the QUBRICS QSO dataset with respect to the Gaia QSO sample is thus equal to 82\% for $z>2.5$.

For all the 2278 QSOs with a spectroscopic redshift, we compared $z_{\rm QU\_G}$, the redshift measured using the Gaia spectrum, with the established spectroscopic redshift $z_{\rm spec}$ known thanks to spectra from SDSS DR16Q \citep[51\%,][]{LykeSDSS16q:2020ApJS..250....8L}, \citet[][19\%]{Veron10:2010A&A...518A..10V}, QUBRICS follow-up campaigns (23\%), \citet[][3\%]{SDSSIncomplete_Schindler:2019ApJ...871..258S} with the remaining fraction from minor sources.

\begin{figure}[h!]
    \centering
    \includegraphics[width=\columnwidth]{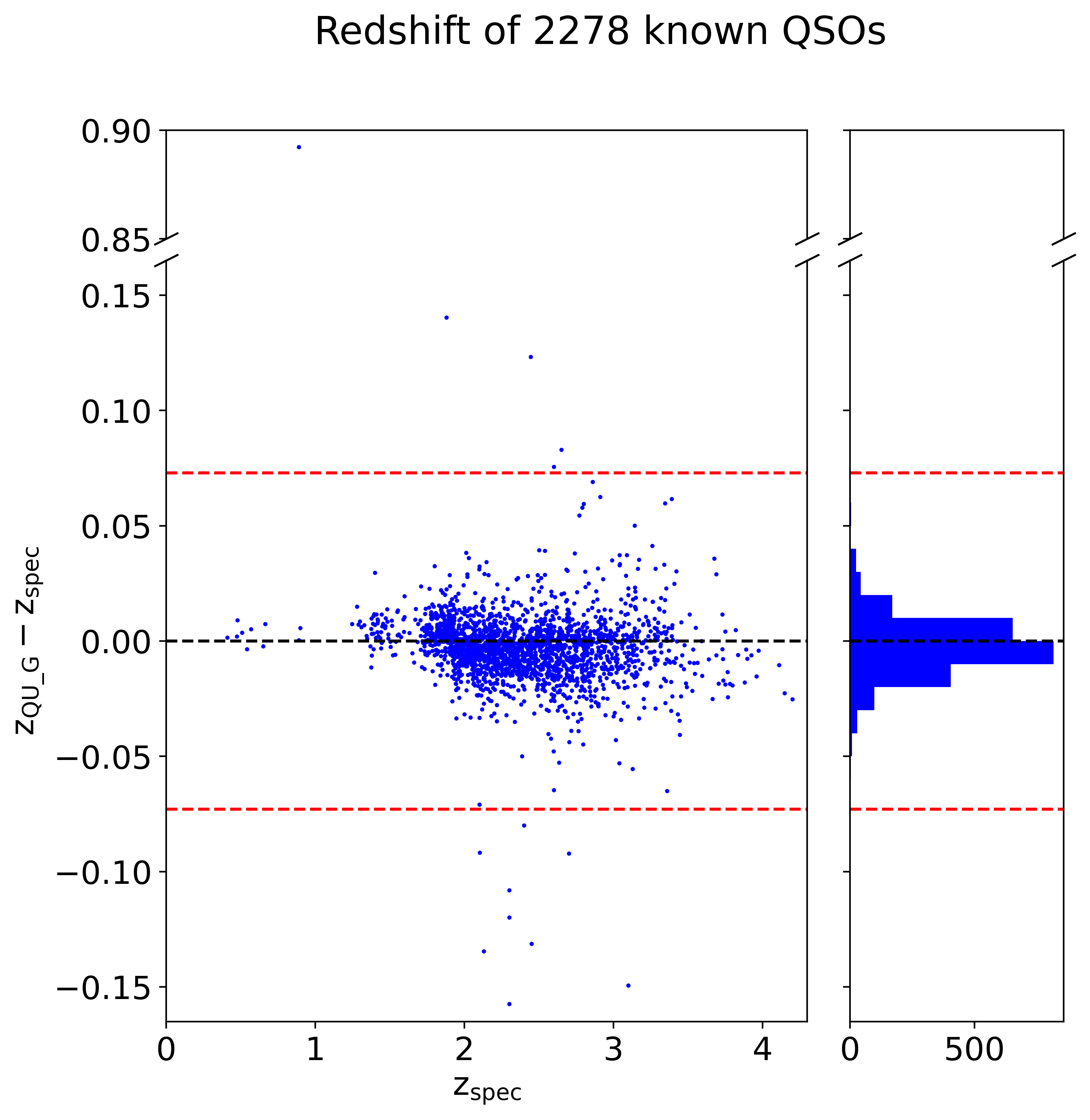}
    \caption{
    The difference between the redshifts determined on the basis of the Gaia low-resolution spectra and the spectroscopic redshifts, $\Delta z$, as a function of the spectroscopic redshift. The dashed red lines mark the $\Delta_z=\pm5\sigma_z$ threshold chosen to identify catastrophic discrepancies.
    }
    \label{Fig:Dz_test_sample}
\end{figure}

The comparison between the measured redshift $z_{\rm QU\_G}$ and the spectroscopic redshift $z_{\rm spec}$ is shown in Fig. \ref{Fig:Dz_test_sample} and shows a good agreement between the two.

The distribution of the differences $\Delta z = |z_{\rm QU\_G}-z_{\rm spec}| $ was used to quantitatively estimate the compatibility between the two measures and to find any discrepancies.
We applied a two-step sigma clipping procedure to identify redshift outliers. First, we computed the standard deviation of all redshift differences $\Delta z = |z_{\mathrm{QU\_G}} - z_{\mathrm{spec}}|$. Objects with $\Delta z>5\sigma$ were temporarily excluded, and the standard deviation was recalculated from the remaining sample, finding $\sigma_{z}=0.015$. This estimate of the standard deviation is robust with respect to outliers.
A threshold of $5 \sigma_z$ was chosen to identify catastrophic discrepancies: 14 objects were found to have $\Delta z > 5 \sigma_z \simeq 0.07$, all of which had a spectroscopic redshift known from the literature:
\begin{itemize}
    \item 11 from the Calan-Tololo Survey \citep{Calan-Tololo:1993RMxAA..25...51M};
    \item 2 from \cite{Veron10:2010A&A...518A..10V};
    \item 1 from SDSS DR16 \citep{LykeSDSS16q:2020ApJS..250....8L}.
\end{itemize}

To further investigate the origin of the redshift discrepancies, we carried out follow-up spectroscopic observations for 12 of the 14 discrepant objects (all those with $DEC<0$; the remaining 2 have $DEC>30$ and were not visible during our observing runs). 
The observational details are provided in Appendix \ref{appendix:discrepancies-spectra}. In all 12 cases, the newly determined spectroscopic redshifts ($z_{\rm spec\_obs}$) were consistent with those derived from the Gaia spectra ($z_{\rm QU\_G}$,  see Table \ref{tab:completeness_outliers_observed}). These results indicate that the discrepancies were due to inaccuracies in the previously published redshifts, rather than errors in the Gaia-based measurements. The comparison with known QSOs and the spectroscopic observations demonstrate that the procedure described in Sect. \ref{sec:define_qso_sample} produces secure redshifts from Gaia low-resolution spectra, with an uncertainty of $\sigma_z \sim 0.015$. For this reason, we can make the following assumption: all the objects in the Gaia QSO sample are QSOs with redshift corresponding to the one measured from the low-resolution spectra (within the measured uncertainty of $\sigma_z \sim 0.015$).

The remaining 1223 QSOs with Gaia spectra, that did not have a known classification or spectroscopic redshift in the QUBRICS QSO dataset, have been added to the QUBRICS database. Most of these new QSOs have low redshift: the median redshift is 2.1, with 16-84 percentile range [1.84, 2.52], and only 205 have $z\ge2.5$. 
The median G magnitude is 17.82, with 16-84 percentile range [17.31, 18.14]. 

A coordinate search of the 1223 new QSOs was performed on the NASA/IPAC Extragalactic Database (NED)\footnote{\url{https://ned.ipac.caltech.edu/}}, using a 0.5 arcsec radius (matching the value adopted for all crossmatches in this work)
to ensure consistency and to minimize false associations. 168 of the 1223 new QSOs already had a published redshift in the literature, mostly from LAMOST \citep{LAMOST-DR4-5:2019ApJS..240....6Y, LAMOST-DR6-9:2023ApJS..265...25J} and DESI \citep{DESI-EDR1:2024AJ....168...58D} surveys, but also from photometric redshift estimates \citep{SDSS-PhotZ:2009ApJS..180...67R}. The large majority of the objects, 1055, do not have a published redshift in the literature and are thus new QSO discoveries enabled by Gaia DR3 spectroscopy. These new objects will also be used to improve the training for future candidate selections.

\section{Estimating the completeness of the XGB selection}
\label{sec:XGB}
We can now proceed to estimate the completeness of the QUBRICS XGB-based selection. As described in Paper II, the XGB algorithm is trained and tested on a set of spectroscopically classified objects, and then is applied to a set of unclassified objects to predict their classification. The set of QSO candidates is composed by the unclassified objects predicted to be high-$z$ QSOs by the XGB selection process.

The XGB dataset starts from the PanSTARRS1 DR2 \citep{PanSTARRS:Chambers_2016} survey and includes all objects having:
\begin{enumerate}
    \item PanSTARRS1DR2 magnitude $y_p$<19;
    \item W1 magnitude (3.4 $\mu$m) from AllWISE or CatWISE;
    \item negligible proper motion ($PM/PM_{err}<3$) from Gaia DR3\footnotemark{};
    \item negligible parallax ($parallax\_over\_error<3$) from Gaia DR3\footnotemark[\value{footnote}];
\end{enumerate}
Unlike Paper II, we did not strictly require a matching source in Gaia DR3, so the dataset is larger by a factor $\sim$2.
\footnotetext{sources with $PM/PM_{err}>3$ or $parallax\_over\_error>3$ from Gaia DR3 were excluded from the dataset, while sources with $PM/PM_{err}\le3$ and $parallax\_over\_error\le3$, or with no matching source in Gaia DR3 were included.}

The composition of the XGB dataset is shown in Tab.~\ref{tab:XGB_dataset}. 
\begin{table}[H]
	\centering
	\caption{Composition of the XGB dataset.}
	\label{tab:XGB_dataset}
    \begin{tabular}{|l | r | r |}
      \hline
          {\bf Class} & {\bf N. of sources} & {\bf Fraction}\\
          \hline
          Stars                  & 64,814,386  &              86.273\% \\
          Galaxies               &     27,729  &               0.037\% \\
          low-$z$ QSOs and AGNs  &     91,453  &               0.122\% \\
          high-$z$ QSOs          &     10,866  &               0.014\% \\
          other\tablefootmark{a} &      4,775  &               0.006\% \\
          {\it unclassified}     &  10,177,696  &               13.547\% \\
          \hline
    \end{tabular}
    \tablefoot{
        \tablefoottext{a}{Any other spectral classification, such as Type 2 AGN, HII region, BL Lac, etc.}
    }
\end{table}
The last line of Tab.~\ref{tab:XGB_dataset} represents the sources for which we could not find a spectroscopic classification, nor significant proper motion or parallax measurements in the Gaia catalog; hence, it is the subset used to search for new high-$z$ ($z>2.5$) QSO candidates.
Of the 3501 QSOs of the sample defined in Sect.\ref{sec:define_qso_sample}, 2713 fall within the footprint of the XGB dataset selection and 894 of them have {$z_{QU\_G} \ge 2.5$}.

We performed a coordinate crossmatch (radius 0.5 arcsec) between the 894 QSOs in our independent sample and the XGB dataset. We found that 18 of these QSOs were missing from the XGB dataset, corresponding to a 2\% shortfall. 
We refer to this shortfall as {\it dataset incompleteness}, which we define as the fraction of QSOs that fall within the survey footprint but are absent from the dataset due to missing photometric or astrometric information.
Specifically:
\begin{itemize}
    \item 1 lacked a PanSTARRS1DR2 match within 0.5";
    \item 4 had no match in AllWISE or CatWISE;
    \item 13 did not have a corresponding $y_p$ magnitude from PanSTARRS1DR2.
\end{itemize}
The remaining 876 QSOs were successfully found in the XGB dataset, either with or without a spectroscopic classification from the QUBRICS QSO dataset.
724 have a spectroscopic classification as QSOs in the dataset, while 152 of the Gaia QSOs were unclassified and should have been selected as QSO candidates by the XGB algorithm. In fact, we find 136 of them in the XGB candidate list. The recall of the XGB candidate selection algorithm is thus \textbf{$136/152=89\%$}, a result in line with the estimate reported in Paper II. If we also account for the XGB dataset completeness (98\%), the overall XGB selection completeness is 87\%.

\begin{figure}
    \centering
    \includegraphics[width=\columnwidth]{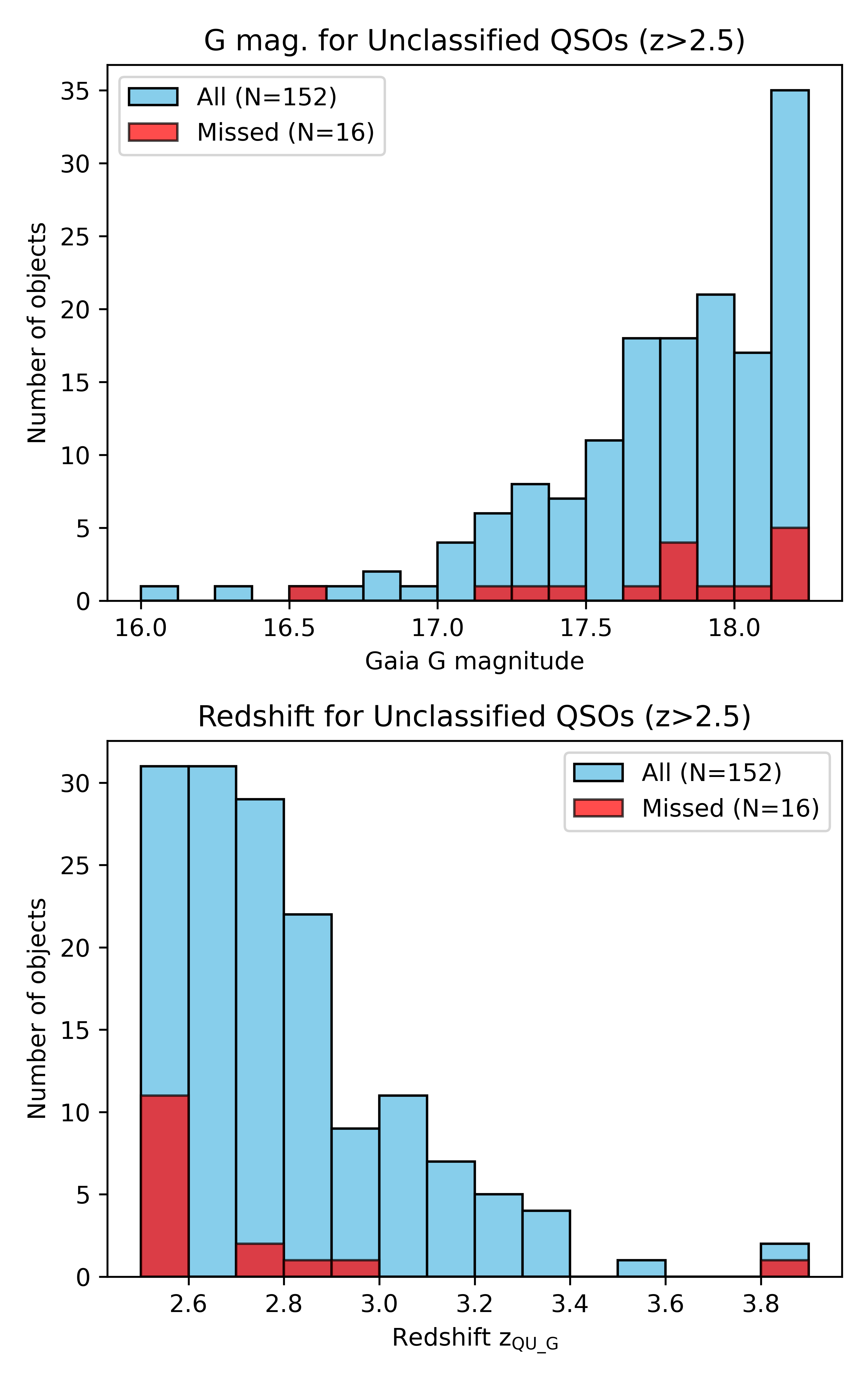}
    \caption{\textbf{Top panel}: histogram of the Gaia G magnitude for the 152 QSOs with no classification in the XGB sample, with the 16 QSOs that were not identified as candidates ("missed") highlighted in red.
    \textbf{Bottom panel}: histogram of the $z_{\rm QU\_G}$ redshifts for the 152 QSOs with no classification in the XGB sample, with the 16 QSOs that were not identified as candidates ("Missed") highlighted in red.}
    \label{Fig:xgb_G-z_hist}
\end{figure}

The 16 QSOs that were not identified as high-redshift QSO candidates were assigned the following predicted categories by the XGB:
\begin{itemize}
    \item 14 low-redshift QSOs (all with $2.5<z_{QU\_G}<2.9$);
    \item 1 star ($z_{QU\_G}=2.95$);
    \item 1 galaxy ($z_{QU\_G}=3.80$);
\end{itemize}
Most missed objects lie close to the $z=2.5$ boundary, where photometric degeneracies with low-redshift QSOs are expected. The distribution in magnitude and redshift of the 152 unclassified QSOs is shown in Fig. \ref{Fig:xgb_G-z_hist}. QSOs that have not been selected as candidates by the XGB are shown in red. The candidate selection recall remains approximately constant across magnitude and redshift, except for the lowest redshift bin: this behavior is expected, due to the lower classification accuracy for objects close to the $z=2.5$ threshold on which the XGB has been trained. In fact, the large majority of the misclassified QSOs had a redshift within 0.4 from the threshold and were predicted to be low-redshift QSOs.

\section{Estimating the completeness of the PRF selection}
\label{sec:PRF}
The second selection method used in the QUBRICS survey is based on the Probabilistic Random Forest algorithm (Paper I). Again, we can use the independent sample defined in Sect. \ref{sec:define_qso_sample} to estimate the completeness and recall of the PRF selection.

The PRF dataset starts from the SkyMapper DR4 survey \citep{SkyMapper4:2024PASA...41...61O} and includes all objects having:
\begin{enumerate}
    \item SkyMapper magnitude $14 < i < 19$;
    \item reliable photometry (quality flag $i\_flags\leq4$);
    \item a W1 magnitude (3.4 $\mu$m) from CatWISE;
    \item negligible proper motion ($PM/PM_{err}<3$) from Gaia DR3\footnotemark{};
    \item negligible parallax ($parallax\_over\_error<3$) from Gaia DR3\footnotemark[\value{footnote}];
\footnotetext{sources with $PM/PM_{err}>3$ or $parallax\_over\_error>3$ from Gaia DR3 were excluded from the dataset, while sources with $PM/PM_{err}\le3$ and $parallax\_over\_error\le3$, or with no matching source in Gaia DR3 were included.}
    \item a PSF-Petrosian magnitude difference smaller than $4\times \sigma\_{\rm ext}$\footnote{$\sigma\_{\rm ext}$ is calculated as the average of the normalized PSF-Petrosian magnitude differences in $i$ and $z$ bands, where each band's difference is normalized by measurement errors and calibrated relative to the median difference for objects of similar magnitude.};
    \item galactic latitude $|b|>15\deg$.
\end{enumerate}

Of the 3501 QSOs of the sample defined in Sect.\ref{sec:define_qso_sample}, 2136 fall within the footprint of the PRF dataset selection and 675 of them have {$z_{QU\_G} \ge 2.5$}. Then, we perform a coordinate crossmatch (radius 0.5") with the PRF dataset, and find that 23 of 675 QSOs in the footprint were not included in the PRF dataset (3\% dataset incompleteness), for the following reasons:
\begin{itemize}
    \item 7 did not have an \textit{i} magnitude from SkyMapper DR4;
    \item 7 had unreliable SkyMapper photometry ($i\_flags>4$);
    \item 9 did not have a W1 magnitude from CatWISE.

\end{itemize}

The remaining 652 QSOs were successfully found in the PRF dataset, either with or without a spectroscopic classification. 69 of these had no known classification and should have been selected as QSO candidates by the PRF algorithm. Checking the candidate list, we find 46 objects out of the 69 sources: this results in a 66\% recall for the PRF candidate selection method. This is slightly lower than the 67.5\% estimate presented in Paper I. If we also account for the PRF dataset completeness (97\%), the overall PRF selection completeness is 64\%.

The 23 QSOs that were not identified as high-redshift QSO candidates were assigned the following predicted categories by the PRF:
\begin{itemize}
    \item 18 low-redshift QSOs (all with $2.5<z_{QU\_G}<2.9$);
    \item 4 stars (with $2.69\le z_{QU\_G}\le3.11$);
    \item 1 galaxy (with $z_{QU\_G}=2.96$);
\end{itemize}

\begin{figure}
    \centering
    \includegraphics[width=\columnwidth]{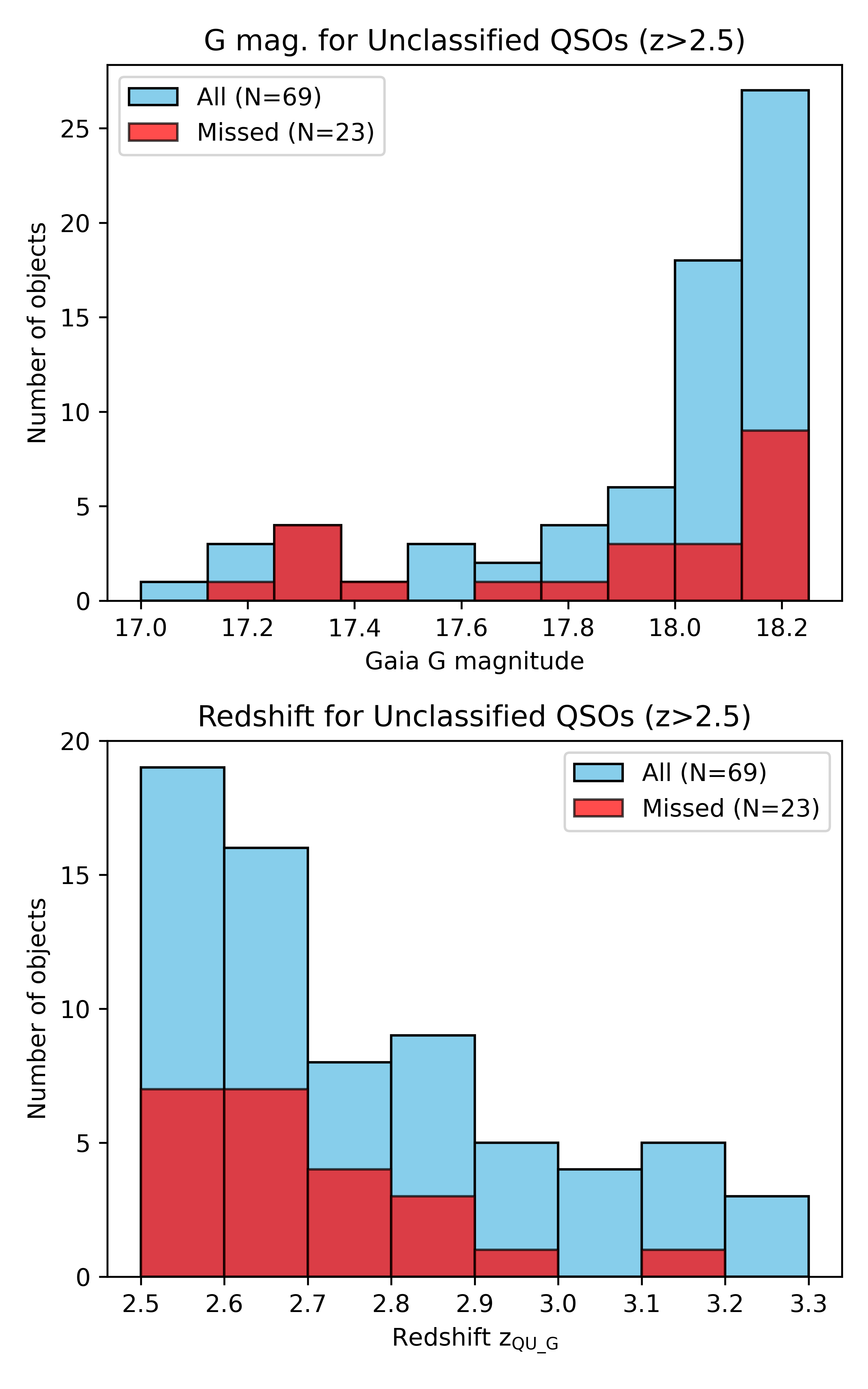}
    \caption{\textbf{Top panel}: histogram of the Gaia G magnitude for the 69 QSOs with no classification in the PRF sample, with the 23 QSOs that were not identified as candidates ("missed") highlighted in red.
    \textbf{Bottom panel}: histogram of the $z_{\rm QU\_G}$ redshifts for the 69 QSOs with no classification in the PRF sample, with the 23 QSOs that were not identified as candidates ("Missed") highlighted in red.}
    \label{Fig:prf_G-z_hist}
\end{figure}

The distribution in magnitude and redshift of the 69 unclassified QSOs is shown in Fig. \ref{Fig:prf_G-z_hist}. QSOs that have not been selected as candidates by the PRF are shown in red. The candidate selection recall is lower for $z<3$; like the XGB case, most of the misclassified QSOs are still predicted to be low-redshift QSOs; however, unlike the XGB case, they are not concentrated in the lowest redshift bin. This is expected because the PRF is strictly a classification algorithm and does not incorporate redshift prediction, while the XGB also performs regression: this reduces the effect given by the threshold at $z>2.5$ in the training set.

\section{Reliability of the recall estimate}
\label{sec:Reliability}
\begin{figure*}[hbtp!]
\sidecaption
    \includegraphics[width=12cm]{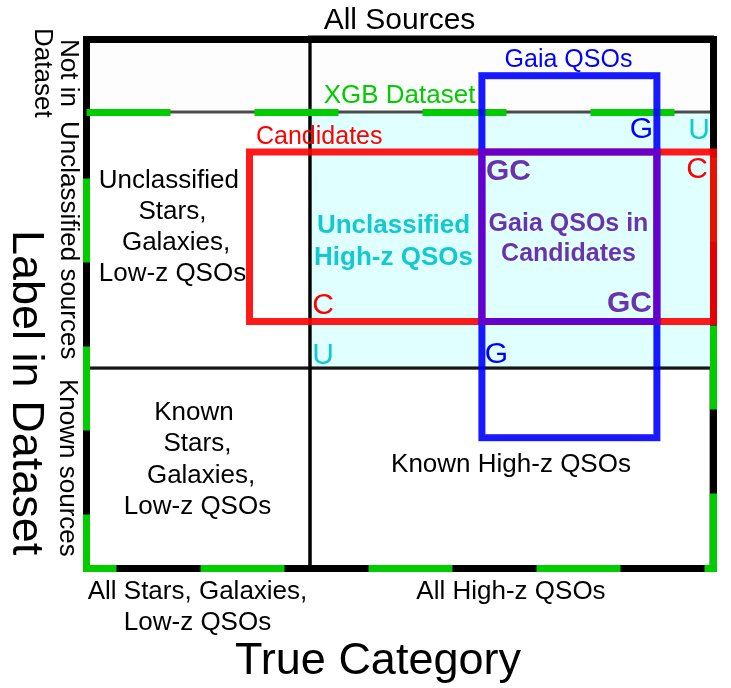}
    \caption{Schematic representation of the datasets used in the QUBRICS survey and in this paper. The outermost black rectangle contains all the sources within a given footprint and magnitude range.
    The green dashed rectangle denotes the dataset used in the XGB selection. Vertical divisions separate the sources according to their true category, while horizontal divisions separate the sources according to their label in the database. Known uninteresting sources (stars, galaxies, low-redshift QSOs) are in the bottom left quadrant; known high-redshift QSOs are in the bottom right quadrant; unclassified sources that are stars, galaxies or low-redshift QSOs are in the top left quadrant; unclassified sources that are high-redshift QSOs are in the top right.
    The region of interest is highlighted in cyan and by the letter {\color{cyan} U}: this is the set of of all true high-redshift QSOs that are unclassified in the dataset. 
    The red rectangle represents the set of QSO candidates predicted by the XGB, and its intersection {\color{red} C} with the {\color{cyan} U} region is the set of unclassified high-redshift QSOs that are also candidates.
    The blue rectangle represents the set of Gaia QSOs, and its intersection {\color{blue} G} with the cyan {\color{cyan} U} region is the set of Gaia QSOs that are unclassified in the dataset.}
    \label{fig:datasets_diagram}  
\sidecaption
    \includegraphics[width=12cm]{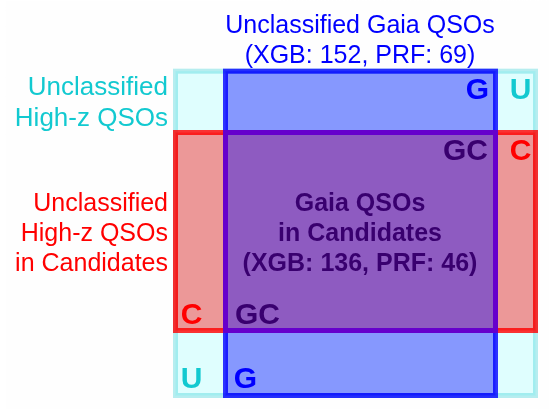}
    \caption{A zoom-in of the region of interest {\color{cyan} U} in Fig. \ref{fig:datasets_diagram}. The intersection {\color{violet} GC} between the blue and red rectangles is the set of unclassified high-redshift QSOs in the dataset that are both QSO candidates and in the Gaia QSO sample.
    Completeness and recall metrics are defined in terms of these intersections in Sect. \ref{sec:Reliability}.}
    \label{fig:datasets_diagram-zoom}  
\end{figure*}

In order to better assess the accuracy of the recall estimates just measured, it is helpful to start from the visual representation of the datasets shown in Fig. \ref{fig:datasets_diagram}.

According to the definition of recall provided in Sect. \ref{sec:define_qso_sample}, the recall $R$ is the fraction:
\begin{equation}
    R = \frac{N_{C}}{N_{U}}
\end{equation}
where $N_{C}$ is the number of unclassified high-redshift QSOs in the dataset that have been selected as candidates and $N_{U}$ is the total number of unclassified high-redshift QSOs in the dataset. 
Both $N_{C}$ and $N_{U}$ are unknown: to discover their value, all unclassified sources would need to be observed to attach a reliable classification.
However, by using the Gaia QSO sample and crossmatching it with the unclassified and candidate sets (see Figs. \ref{fig:datasets_diagram} and \ref{fig:datasets_diagram-zoom}), we can obtain the recall estimate $R_G$ with:
\begin{equation}
    R_G = \frac{N_{GC}}{N_{G}}
\end{equation}
where $N_{GC}$ is the number of Gaia QSO that have been selected as candidates and $N_{G}$ is the total number of Gaia QSOs that are unclassified in the dataset. 
The recall estimates $R_G$ obtained in this way were 89\% for the XGB and 66\% for the PRF selection (see Sects. \ref{sec:XGB} and \ref{sec:PRF}). We can derive their reliability with respect to the true recall $R$ by making the following assumptions:
\begin{enumerate}
    \item each of the $N_{G}$ unclassified Gaia QSOs is picked randomly among the unclassified QSOs;
    \item $N_{G}\ll N_{U}$, so that the probability of picking each of the $N_{G}$ QSOs remains approximately constant.
\end{enumerate}
It follows from these assumptions that each of the $N_{G}$ unclassified Gaia QSOs has a probability $R$ of being among the candidates, and a probability $1-R$ of not being a candidate.
Thus, the number $N_{GC}$ of unclassified Gaia QSOs that are also QSO candidates follows a binomial distribution, with expected value and variance:
\begin{equation}
    E[N_{GC}] = N_{G} \times R
\end{equation}
\begin{equation}
    {\it Var}[N_{GC}] = N_{G} \times R(1-R)
\end{equation}
By considering the quantity $R_G=N_{GC}/N_{G}$, we obtain:
\begin{equation}
\label{eq:expR_G}
    E[R_G] = R
\end{equation}
\begin{equation}
\label{eq:stdR_G}
    {\it Var}[R_G] = \frac{R(1-R)}{N_{G}}
\end{equation}
Using the numbers $N_{GC}$ and $N_{G}$ obtained in Sect.~\ref{sec:XGB} and \ref{sec:PRF} we can estimate $R_G$ and according to Eq.~\ref{eq:expR_G} this is also an estimate for $R$.  Similarly, we can estimate an order-of-magnitude uncertainty for $R$ by replacing $R$ with $R_{G}$ in Eq.~\ref{eq:stdR_G}.  For the XGB selection (Sect.~\ref{sec:XGB}) we have $N_{GC}=136$ and $N_{G}=152$, hence:
\begin{equation}
    R = 0.89 \pm \sqrt{\frac{0.89(1-0.89)}{152}} = 0.89 \pm 0.03,
\end{equation}
while for the PRF selection we have (Sect.~\ref{sec:PRF}), $N_{GC}=46$ and $N_{G}=69$, hence:
\begin{equation}
    R = 0.67 \pm \sqrt{\frac{0.67(1-0.67)}{69}} = 0.67 \pm 0.06.
\end{equation}

These values are valid as long as our assumptions are valid. 
However, since the number of unclassified QSOs $N_{U}$ is unknown, we should not assume a priori that $N_{G}\ll N_{U}$ and the case $N_{G} < N_{U}$ should also be considered. If a significant fraction of the $N_{U}$ unclassified QSOs is contained in the Gaia QSO sample, then $N_{GC}$ deviates from a binomial distribution. Nonetheless, these values are still useful as an order-of-magnitude estimate of the uncertainty.

\section{Conclusions}
\label{sec:conclusions}

In this work, we have presented an independent assessment of the completeness and recall of the QUBRICS QSO selection algorithms - XGB \citep{Calderone2024} and PRF \citep{Guarneri2022} - using a robust sample of 3501 QSOs identified from Gaia DR3 low-resolution spectra. 
This analysis is essential for quantifying the reliability of QUBRICS selections, in view of a wide range of cosmological applications, including studies of the QSO luminosity function, cosmic reionization, and the redshift drift.
We have obtained the following benchmarks:
\begin{enumerate}
    \item a measure of the spectroscopic completeness (number of spectroscopically confirmed QSOs in the dataset divided by the number of true QSOs in the sky) of the QUBRICS QSO dataset with respect to the Gaia QSO sample: we found in Sect. \ref{sec:validate_qso_sample} that 82\% of the QSOs in the Gaia sample with $z>2.5$ were already classified as QSOs in the QUBRICS dataset of spectroscopically classified sources;
    \item estimates of the dataset completeness (number of true QSOs in the dataset divided by the number of true QSOs in the sky) for the XGB and PRF datasets: we found that 97-98\% of the Gaia QSOs in datasets footprints were also present in the datasets, either as classified or unclassified objects, a reasonably high value; 
    \item estimates of the recall (number of predicted QSOs that are actually QSOs divided by the number of true QSOs that have no classification in the dataset) for the two main selection algorithms, XGB and PRF: we found that the XGB algorithm correctly identified 89\% of the unclassified Gaia QSOs in its dataset, while the PRF algorithm correctly identified 66\% of the unclassified Gaia QSOs in its dataset;
    \item estimates of the selection completeness (number of predicted QSOs that are actually QSOs divided by the number of true QSOs in the dataset footprint without a classification): combining the dataset completeness and recall estimates, we found that the XGB-selected candidates are 87\% of the unclassified Gaia QSOs in the PanSTARRS1 footprint, while the PRF-selected candidates are 64\% of the unclassified Gaia QSOs in the SkyMapper footprint.
\end{enumerate}

Taken together, these results paint the overall picture of the QUBRICS survey: the completeness of spectroscopically confirmed QSOs is currently around 82\%, and observing the QSO candidates provided by the XGB and PRF will increase it to $\sim$~87\% (the selection completeness).

We remark that the reliability of the recall estimates depends on the assumptions described in Sec. \ref{sec:Reliability}. The completeness and recall estimates derived here were computed using objects within the redshift ranges $ 2.5 < z < 3.4 $ for XGB and $ 2.5 < z < 3.3 $ for PRF (see Figs. \ref{Fig:xgb_G-z_hist} and \ref{Fig:prf_G-z_hist}). 
These ranges are set by the limited number of QSOs with $z>3.5$ identified in the Gaia sample. Our empirical validation therefore applies mainly over the interval $2.5<z<3.4$ (with the precise upper limit differing slightly between the XGB and PRF selections). Extrapolation beyond this redshift range should be treated with caution; however, we find no indication within our data that the selection performance changes abruptly at higher redshift.
The values for the recall are consistent with previous internal estimates \citep[86\% for XGB and 68\% for PRF;][]{Calderone2024, Guarneri2022} but now benefit from validation against an independent dataset, strengthening their credibility.

The difference in performance between the two algorithms can be attributed to their methodological foundations. XGB, which incorporates both classification and regression, is better suited to handling imbalanced datasets and predicting redshifts, especially near the critical threshold of $z \sim 2.5$. In contrast, PRF is a pure classification algorithm and lacks the ability to model redshift-dependent selection effects, which likely contributes to its reduced recall. 
Future improvements in recall will likely require expanded training sets and integration of additional photometric and spectroscopic surveys, such as \textit{Vera Rubin Observatory / Legacy Survey of Space and Time (VRO/LSST)} \citep{LSST_guy_2025_15558559}, \textit{Euclid} \citep{Euclid:2022A&A...662A.112E, Euclid_2025A&A...697A...1E} and the \textit{Nancy Grace Roman Space Telescope} \citep{RomanSpaceTelecope:2022ApJ...928....1W}.

Our analysis also led to the identification of 1223 new QSOs, including 205 with $ z > 2.5$, that were previously unclassified. These objects not only enrich the QUBRICS database but also provide a valuable training set for future iterations of the selection algorithms. The validation of Gaia-derived redshifts through targeted follow-up spectroscopy confirms the reliability of the independent sample, with a redshift uncertainty of 
$\sigma_z \sim 0.015$.

From a cosmological perspective, the high recall of the XGB method ensures that QUBRICS-selected QSOs can be reliably used for statistical studies, such as measuring the QSO luminosity function and probing the IGM. The identification of missed candidates by both algorithms also highlights areas for methodological refinement, particularly in the treatment of photometric uncertainties and the inclusion of synthetic data.

In conclusion, by benchmarking the performance of the XGB and PRF algorithms against an external sample, we obtain a better understanding of their strengths and limitations. These insights will guide the refinement of future selection strategies and improve the scientific utility of QUBRICS, particularly in preparation for upcoming observational facilities such as 40m-class telescopes.

   \bibliographystyle{aa}
   \bibliography{References}

@ARTICLE{Trost2025,
	author = {{Trost, A.} and {Marques, C. M. J.} and {Cristiani, S.} and {Cupani, G.} and {Di Stefano, S.} and {D\'{}Odorico, V.} and {Guarneri, F.} and {Martins, C. J. A. P.} and {Milakovi\'{}c, D.} and {Pasquini, L.} and {G\'enova Santos, R.} and {Molaro, P.} and {Murphy, M. T.} and {Nunes, N. J.} and {Schmidt, T. M.} and {Alibert, Y.} and {Boutsia, K.} and {Calderone, G.} and {Gonz\'alez Hern\'andez, J. I.} and {Grazian, A.} and {Lo Curto, G.} and {Palle, E.} and {Pepe, F.} and {Porru, M.} and {Santos, N. C.} and {Sozzetti, A.} and {Su\'arez Mascare\~no, A.} and {Zapatero Osorio, M. R.}},
	title = {The ESPRESSO Redshift Drift Experiment - I. High-resolution spectra of the Lyman-est of QSO J052915.80-435152.0},
	DOI= "10.1051/0004-6361/202554502",
	url= "https://doi.org/10.1051/0004-6361/202554502",
	journal = {A\&A},
	year = 2025,
	volume = 699,
	pages = "A159",
}

@ARTICLE{Trost2026,
       author = {{Trost}, Andrea and {Marques}, Catarina M.~J. and {Cristiani}, S. and {Cupani}, Guido and {Di Stefano}, Simona and {D'Odorico}, Valentina and {Guarneri}, Francesco and {Martins}, Carlos J.~A.~P. and {Milakovi{\'c}}, Dinko and {Pasquini}, Luca and {G{\'e}nova Santos}, Ricardo and {Molaro}, Paolo and {Murphy}, Michael T. and {Nunes}, Nelson J. and {Schmidt}, Tobias M. and {Alibert}, Yann and {Boutsia}, Konstantina and {Calderone}, Giorgio and {Gonz{\'a}lez Hern{\'a}ndez}, J.~I. and {Grazian}, Andrea and {Lo Curto}, Gaspare and {Palle}, Enric and {Pepe}, Francesco and {Porru}, Matteo and {Santos}, Nuno C. and {Sozzetti}, Alessandro and {Su{\'a}rez Mascare{\~n}o}, Alejandro and {Zapatero Osorio}, Maria R.},
        title = "{The ESPRESSO Redshift Drift Experiment III -- The Third Epoch of QSO J052915.80-435152.0}",
      journal = {arXiv e-prints},
         year = 2026,
        month = mar,
          eid = {arXiv:2603.02318},
        pages = {arXiv:2603.02318},
archivePrefix = {arXiv},
       eprint = {2603.02318},
 primaryClass = {astro-ph.CO},
       adsurl = {https://ui.adsabs.harvard.edu/abs/2026arXiv260302318T}
}

@ARTICLE{Madau2024,
       author = {{Madau}, Piero and {Giallongo}, Emanuele and {Grazian}, Andrea and {Haardt}, Francesco},
        title = "{Cosmic Reionization in the JWST Era: Back to AGNs?}",
      journal = {\apj},
         year = 2024,
        month = aug,
       volume = {971},
       number = {1},
          eid = {75},
        pages = {75},
          doi = {10.3847/1538-4357/ad5ce8},
archivePrefix = {arXiv},
       eprint = {2406.18697},
 primaryClass = {astro-ph.CO},
       adsurl = {https://ui.adsabs.harvard.edu/abs/2024ApJ...971...75M}
}

@ARTICLE{Grazian2022,
       author = {{Grazian}, Andrea and {Giallongo}, Emanuele and {Boutsia}, Konstantina and {Calderone}, Giorgio and {Cristiani}, Stefano and {Cupani}, Guido and {Fontanot}, Fabio and {Guarneri}, Francesco and {Ozdalkiran}, Yacob},
        title = "{The Space Density of Ultra-luminous QSOs at the End of Reionization Epoch by the QUBRICS Survey and the AGN Contribution to the Hydrogen Ionizing Background}",
      journal = {\apj},
         year = 2022,
        month = jan,
       volume = {924},
       number = {2},
          eid = {62},
        pages = {62},
          doi = {10.3847/1538-4357/ac33a4},
archivePrefix = {arXiv},
       eprint = {2110.13736},
 primaryClass = {astro-ph.GA},
       adsurl = {https://ui.adsabs.harvard.edu/abs/2022ApJ...924...62G}
}

@INPROCEEDINGS{MIDAS1988,
       author = {{Banse}, K. and {Ponz}, D. and {Ounnas}, C. and {Grosbol}, P. and {Warmels}, R.},
        title = "{The MIDAS Image Processing System}",
    booktitle = {Instrumentation for Ground-Based Optical Astronomy},
         year = 1988,
        month = jan,
        pages = {431},
       adsurl = {https://ui.adsabs.harvard.edu/abs/1988igbo.conf..431B}
}

@inproceedings{GOODMAN2004,
author = {J. Christopher Clemens and J. Adam Crain and Robert Anderson},
title = {{The Goodman spectrograph}},
volume = {5492},
booktitle = {Ground-based Instrumentation for Astronomy},
editor = {Alan F. M. Moorwood and Masanori Iye},
organization = {International Society for Optics and Photonics},
publisher = {SPIE},
pages = {331 -- 340},
year = {2004},
doi = {10.1117/12.550069},
URL = {https://doi.org/10.1117/12.550069}
}

@INPROCEEDINGS{GOODMAN_Pipeline2020,
       author = {{Torres-Robledo}, Sim{\'o}n and {Brice{\~n}o}, C{\'e}sar and {Quint}, Bruno and {Sanmartim}, David},
        title = "{Spectroscopic Data Reduction Pipeline for the Goodman High Throughput Spectrograph}",
    booktitle = {Astronomical Data Analysis Software and Systems XXVII},
         year = 2020,
       editor = {{Ballester}, Pascal and {Ibsen}, Jorge and {Solar}, Mauricio and {Shortridge}, Keith},
       series = {Astronomical Society of the Pacific Conference Series},
       volume = {522},
        month = apr,
        pages = {533},
       adsurl = {https://ui.adsabs.harvard.edu/abs/2020ASPC..522..533T}
}

@ARTICLE{Bosman2022,
       author = {{Bosman}, Sarah E.~I. and {Davies}, Frederick B. and {Becker}, George D. and {Keating}, Laura C. and {Davies}, Rebecca L. and {Zhu}, Yongda and {Eilers}, Anna-Christina and {D'Odorico}, Valentina and {Bian}, Fuyan and {Bischetti}, Manuela and {Cristiani}, Stefano V. and {Fan}, Xiaohui and {Farina}, Emanuele P. and {Haehnelt}, Martin G. and {Hennawi}, Joseph F. and {Kulkarni}, Girish and {Mesinger}, Andrei and {Meyer}, Romain A. and {Onoue}, Masafusa and {Pallottini}, Andrea and {Qin}, Yuxiang and {Ryan-Weber}, Emma and {Schindler}, Jan-Torge and {Walter}, Fabian and {Wang}, Feige and {Yang}, Jinyi},
        title = "{Hydrogen reionization ends by z = 5.3: Lyman-{\ensuremath{\alpha}} optical depth measured by the XQR-30 sample}",
      journal = {\mnras},
         year = 2022,
        month = jul,
       volume = {514},
       number = {1},
        pages = {55-76},
          doi = {10.1093/mnras/stac1046},
archivePrefix = {arXiv},
       eprint = {2108.03699},
 primaryClass = {astro-ph.CO},
       adsurl = {https://ui.adsabs.harvard.edu/abs/2022MNRAS.514...55B}
}

@ARTICLE{Boutsia2020,
       author = {{Boutsia}, Konstantina and {Grazian}, Andrea and {Calderone}, Giorgio and {Cristiani}, Stefano and {Cupani}, Guido and {Guarneri}, Francesco and {Fontanot}, Fabio and {Amorin}, Ricardo and {D'Odorico}, Valentina and {Giallongo}, Emanuele and {Salvato}, Mara and {Omizzolo}, Alessandro and {Romano}, Michael and {Menci}, Nicola},
        title = "{The Spectroscopic Follow-up of the QUBRICS Bright Quasar Survey}",
      journal = {\apjs},
         year = 2020,
        month = oct,
       volume = {250},
       number = {2},
          eid = {26},
        pages = {26},
          doi = {10.3847/1538-4365/abafc1},
archivePrefix = {arXiv},
       eprint = {2008.03865},
 primaryClass = {astro-ph.GA},
       adsurl = {https://ui.adsabs.harvard.edu/abs/2020ApJS..250...26B}
}

@ARTICLE{Calderone2024,
       author = {{Calderone}, Giorgio and {Guarneri}, Francesco and {Porru}, Matteo and {Cristiani}, Stefano and {Grazian}, Andrea and {Nicastro}, Luciano and {Bischetti}, Manuela and {Boutsia}, Konstantina and {Cupani}, Guido and {D'Odorico}, Valentina and {Feruglio}, Chiara and {Fontanot}, Fabio},
        title = "{Boost recall in quasi-stellar object selection from highly imbalanced photometric datasets. The reverse selection method}",
      journal = {\aap},
         year = 2024,
        month = mar,
       volume = {683},
          eid = {A34},
        pages = {A34},
          doi = {10.1051/0004-6361/202346625},
archivePrefix = {arXiv},
       eprint = {2312.13194},
 primaryClass = {astro-ph.IM},
       adsurl = {https://ui.adsabs.harvard.edu/abs/2024A&A...683A..34C}
}

@ARTICLE{Cristiani2023,
       author = {{Cristiani}, Stefano and {Porru}, Matteo and {Guarneri}, Francesco and {Calderone}, Giorgio and {Boutsia}, Konstantina and {Grazian}, Andrea and {Cupani}, Guido and {D'Odorico}, Valentina and {Fontanot}, Fabio and {Martins}, Carlos J.~A.~P. and {Marques}, Catarina M.~J. and {Maitra}, Soumak and {Trost}, Andrea},
        title = "{Spectroscopy of QUBRICS quasar candidates: 1672 new redshifts and a golden sample for the Sandage test of the redshift drift}",
      journal = {\mnras},
         year = 2023,
        month = jun,
       volume = {522},
       number = {2},
        pages = {2019-2028},
          doi = {10.1093/mnras/stad1007},
archivePrefix = {arXiv},
       eprint = {2304.00362},
 primaryClass = {astro-ph.CO},
       adsurl = {https://ui.adsabs.harvard.edu/abs/2023MNRAS.522.2019C}
}

@ARTICLE{Dayal2024,
	author = {{Dayal, Pratika} and {Volonteri, Marta} and {Greene, Jenny E.} and {Kokorev, Vasily} and {Goulding, Andy D.} and {Williams, Christina C.} and {Furtak, Lukas J.} and {Zitrin, Adi} and {Atek, Hakim} and {Bezanson, Rachel} and {Chemerynska, Iryna} and {Feldmann, Robert} and {Glazebrook, Karl} and {Labbe, Ivo} and {Nanayakkara, Themiya} and {Oesch, Pascal A.} and {Weaver, John R.}},
	title = {UNCOVERing the contribution of black holes to reionization},
	DOI= "10.1051/0004-6361/202449331",
	url= "https://doi.org/10.1051/0004-6361/202449331",
	journal = {A\&A},
	year = 2025,
	volume = 697,
	pages = "A211",
}

@ARTICLE{Fontanot2023,
       author = {{Fontanot}, Fabio and {Cristiani}, Stefano and {Grazian}, Andrea and {Haardt}, Francesco and {D'Odorico}, Valentina and {Boutsia}, Konstantina and {Calderone}, Giorgio and {Cupani}, Guido and {Guarneri}, Francesco and {Fiorin}, Chiara and {Rodighiero}, Giulia},
        title = "{Eddington accreting black holes in the epoch of reionization}",
      journal = {\mnras},
         year = 2023,
        month = mar,
       volume = {520},
       number = {1},
        pages = {740-749},
          doi = {10.1093/mnras/stad189},
archivePrefix = {arXiv},
       eprint = {2301.07129},
 primaryClass = {astro-ph.CO},
       adsurl = {https://ui.adsabs.harvard.edu/abs/2023MNRAS.520..740F}
}

@ARTICLE{LykeSDSS16q:2020ApJS..250....8L,
       author = {{Lyke}, Brad W. and {Higley}, Alexandra N. and {McLane}, J.~N. and {Schurhammer}, Danielle P. and {Myers}, Adam D. and {Ross}, Ashley J. and {Dawson}, Kyle and {Chabanier}, Sol{\`e}ne and {Martini}, Paul and {Busca}, Nicol{\'a}s G. and {Mas des Bourboux}, H{\'e}lion du and {Salvato}, Mara and {Streblyanska}, Alina and {Zarrouk}, Pauline and {Burtin}, Etienne and {Anderson}, Scott F. and {Bautista}, Julian and {Bizyaev}, Dmitry and {Brandt}, W.~N. and {Brinkmann}, Jonathan and {Brownstein}, Joel R. and {Comparat}, Johan and {Green}, Paul and {de la Macorra}, Axel and {Mu{\~n}oz Guti{\'e}rrez}, Andrea and {Hou}, Jiamin and {Newman}, Jeffrey A. and {Palanque-Delabrouille}, Nathalie and {P{\^a}ris}, Isabelle and {Percival}, Will J. and {Petitjean}, Patrick and {Rich}, James and {Rossi}, Graziano and {Schneider}, Donald P. and {Smith}, Alexander and {Vivek}, M. and {Weaver}, Benjamin Alan},
        title = "{The Sloan Digital Sky Survey Quasar Catalog: Sixteenth Data Release}",
      journal = {\apjs},
         year = 2020,
        month = sep,
       volume = {250},
       number = {1},
          eid = {8},
        pages = {8},
          doi = {10.3847/1538-4365/aba623},
archivePrefix = {arXiv},
       eprint = {2007.09001},
 primaryClass = {astro-ph.GA},
       adsurl = {https://ui.adsabs.harvard.edu/abs/2020ApJS..250....8L}
}

@ARTICLE{MARZ_2016,
       author = {{Hinton}, S.~R. and {Davis}, Tamara M. and {Lidman}, C. and {Glazebrook}, K. and {Lewis}, G.~F.},
        title = "{MARZ: Manual and automatic redshifting software}",
      journal = {Astronomy and Computing},
         year = 2016,
        month = apr,
       volume = {15},
        pages = {61-71},
          doi = {10.1016/j.ascom.2016.03.001},
archivePrefix = {arXiv},
       eprint = {1603.09438},
 primaryClass = {astro-ph.IM},
       adsurl = {https://ui.adsabs.harvard.edu/abs/2016A&C....15...61H}
}

@ARTICLE{McQuinn2016,
       author = {{McQuinn}, Matthew},
        title = "{The Evolution of the Intergalactic Medium}",
      journal = {\araa},
         year = 2016,
        month = sep,
       volume = {54},
        pages = {313-362},
          doi = {10.1146/annurev-astro-082214-122355},
archivePrefix = {arXiv},
       eprint = {1512.00086},
 primaryClass = {astro-ph.CO},
       adsurl = {https://ui.adsabs.harvard.edu/abs/2016ARA&A..54..313M}
}

@ARTICLE{Meiksin2009,
       author = {{Meiksin}, Avery A.},
        title = "{The physics of the intergalactic medium}",
      journal = {Reviews of Modern Physics},
         year = 2009,
        month = oct,
       volume = {81},
       number = {4},
        pages = {1405-1469},
          doi = {10.1103/RevModPhys.81.1405},
archivePrefix = {arXiv},
       eprint = {0711.3358},
 primaryClass = {astro-ph},
       adsurl = {https://ui.adsabs.harvard.edu/abs/2009RvMP...81.1405M}
}

@ARTICLE{Murphy2022,
       author = {{Murphy}, Michael T. and {Molaro}, Paolo and {Leite}, Ana C.~O. and {Cupani}, Guido and {Cristiani}, Stefano and {D'Odorico}, Valentina and {G{\'e}nova Santos}, Ricardo and {Martins}, Carlos J.~A.~P. and {Milakovi{\'c}}, Dinko and {Nunes}, Nelson J. and {Schmidt}, Tobias M. and {Pepe}, Francesco A. and {Rebolo}, Rafael and {Santos}, Nuno C. and {Sousa}, S{\'e}rgio G. and {Zapatero Osorio}, Maria-Rosa and {Amate}, Manuel and {Adibekyan}, Vardan and {Alibert}, Yann and {Allende Prieto}, Carlos and {Baldini}, Veronica and {Benz}, Willy and {Bouchy}, Fran{\c{c}}ois and {Cabral}, Alexandre and {Dekker}, Hans and {Di Marcantonio}, Paolo and {Ehrenreich}, David and {Figueira}, Pedro and {Gonz{\'a}lez Hern{\'a}ndez}, Jonay I. and {Landoni}, Marco and {Lovis}, Christophe and {Lo Curto}, Gaspare and {Manescau}, Antonio and {M{\'e}gevand}, Denis and {Mehner}, Andrea and {Micela}, Giuseppina and {Pasquini}, Luca and {Poretti}, Ennio and {Riva}, Marco and {Sozzetti}, Alessandro and {Mascare{\~n}o}, Alejandro Su{\'a}rez and {Udry}, St{\'e}phane and {Zerbi}, Filippo},
        title = "{Fundamental physics with ESPRESSO: Precise limit on variations in the fine-structure constant towards the bright quasar HE 0515{\ensuremath{-}}4414}",
      journal = {\aap},
         year = 2022,
        month = feb,
       volume = {658},
          eid = {A123},
        pages = {A123},
          doi = {10.1051/0004-6361/202142257},
archivePrefix = {arXiv},
       eprint = {2112.05819},
 primaryClass = {astro-ph.CO},
       adsurl = {https://ui.adsabs.harvard.edu/abs/2022A&A...658A.123M}
}

@ARTICLE{SDSSIncomplete_Schindler:2019ApJ...871..258S,
       author = {{Schindler}, Jan-Torge and {Fan}, Xiaohui and {McGreer}, Ian D. and {Yang}, Jinyi and {Wang}, Feige and {Green}, Richard and {Fynbo}, Johan P.~U. and {Krogager}, Jens-Kristian and {Green}, Elisabeth M. and {Huang}, Yun-Hsin and {Kadowaki}, Jennifer and {Patej}, Anna and {Wu}, Ya-Lin and {Yue}, Minghao},
        title = "{The Extremely Luminous Quasar Survey in the Sloan Digital Sky Survey Footprint. III. The South Galactic Cap Sample and the Quasar Luminosity Function at Cosmic Noon}",
      journal = {\apj},
         year = 2019,
        month = feb,
       volume = {871},
       number = {2},
          eid = {258},
        pages = {258},
          doi = {10.3847/1538-4357/aaf86c},
archivePrefix = {arXiv},
       eprint = {1812.04639},
 primaryClass = {astro-ph.GA},
       adsurl = {https://ui.adsabs.harvard.edu/abs/2019ApJ...871..258S}
}

@ARTICLE{LF_Boutsia:2021ApJ...912..111B,
       author = {{Boutsia}, Konstantina and {Grazian}, Andrea and {Fontanot}, Fabio and {Giallongo}, Emanuele and {Menci}, Nicola and {Calderone}, Giorgio and {Cristiani}, Stefano and {D'Odorico}, Valentina and {Cupani}, Guido and {Guarneri}, Francesco and {Omizzolo}, Alessandro},
        title = "{The Luminosity Function of Bright QSOs at z {\ensuremath{\sim}} 4 and Implications for the Cosmic Ionizing Background}",
      journal = {\apj},
         year = 2021,
        month = may,
       volume = {912},
       number = {2},
          eid = {111},
        pages = {111},
          doi = {10.3847/1538-4357/abedb5},
archivePrefix = {arXiv},
       eprint = {2103.10446},
 primaryClass = {astro-ph.GA},
       adsurl = {https://ui.adsabs.harvard.edu/abs/2021ApJ...912..111B}
}

@ARTICLE{Guarneri:2021MNRAS.506.2471G,
       author = {{Guarneri}, Francesco and {Calderone}, Giorgio and {Cristiani}, Stefano and {Fontanot}, Fabio and {Boutsia}, Konstantina and {Cupani}, Guido and {Grazian}, Andrea and {D'Odorico}, Valentina},
        title = "{The probabilistic random forest applied to the selection of quasar candidates in the QUBRICS survey}",
      journal = {\mnras},
         year = 2021,
        month = sep,
       volume = {506},
       number = {2},
        pages = {2471-2481},
          doi = {10.1093/mnras/stab1867},
archivePrefix = {arXiv},
       eprint = {2106.12990},
 primaryClass = {astro-ph.IM},
       adsurl = {https://ui.adsabs.harvard.edu/abs/2021MNRAS.506.2471G}
}

@ARTICLE{Guarneri2022,
       author = {{Guarneri}, Francesco and {Calderone}, Giorgio and {Cristiani}, Stefano and {Porru}, Matteo and {Fontanot}, Fabio and {Boutsia}, Konstantina and {Cupani}, Guido and {Grazian}, Andrea and {D'Odorico}, Valentina and {Murphy}, Michael T. and {Bongiorno}, Angela and {Saccheo}, Ivano and {Nicastro}, Luciano},
        title = "{The probabilistic random forest applied to the QUBRICS survey: improving the selection of high-redshift quasars with synthetic data}",
      journal = {\mnras},
         year = 2022,
        month = dec,
       volume = {517},
       number = {2},
        pages = {2436-2453},
          doi = {10.1093/mnras/stac2733},
archivePrefix = {arXiv},
       eprint = {2209.07257},
 primaryClass = {astro-ph.IM},
       adsurl = {https://ui.adsabs.harvard.edu/abs/2022MNRAS.517.2436G}
}

@ARTICLE{Calderone19:2019ApJ...887..268C,
       author = {{Calderone}, Giorgio and {Boutsia}, Konstantina and {Cristiani}, Stefano and {Grazian}, Andrea and {Amorin}, Ricardo and {D'Odorico}, Valentina and {Cupani}, Guido and {Fontanot}, Fabio and {Salvato}, Mara},
        title = "{Finding the Brightest Cosmic Beacons in the Southern Hemisphere}",
      journal = {\apj},
         year = 2019,
        month = dec,
       volume = {887},
       number = {2},
          eid = {268},
        pages = {268},
          doi = {10.3847/1538-4357/ab510a},
archivePrefix = {arXiv},
       eprint = {1909.06391},
 primaryClass = {astro-ph.IM},
       adsurl = {https://ui.adsabs.harvard.edu/abs/2019ApJ...887..268C}
}

@inproceedings{XGBoost2016,
 author = {Chen, Tianqi and Guestrin, Carlos},
 title = {{XGBoost}: A Scalable Tree Boosting System},
 booktitle = {Proceedings of the 22nd ACM SIGKDD International Conference on Knowledge Discovery and Data Mining},
 series = {KDD '16},
 year = {2016},
 isbn = {978-1-4503-4232-2},
 location = {San Francisco, California, USA},
 pages = {785--794},
 numpages = {10},
 url = {http://doi.acm.org/10.1145/2939672.2939785},
 doi = {10.1145/2939672.2939785},
 acmid = {2939785},
 publisher = {ACM},
 address = {New York, NY, USA},
}

@INPROCEEDINGS{ANDES_2024,
       author = {{Marconi}, A. and {Abreu}, M. and {Adibekyan}, V. and {Alberti}, V. and {Albrecht}, S. and {Alcaniz}, J. and {Aliverti}, M. and {Allende Prieto}, C. and {Alvarado-Gomez}, J.~D. and {Alves}, C.~S. and {Amado}, P.~J. and {Amate}, M. and {Andersen}, M.~I. and {Antoniucci}, S. and {Artigau}, E. and {Bailet}, C. and {Baker}, C. and {Baldini}, V. and {Balestra}, A. and {Barnes}, S.~A. and {Baron}, F. and {Barros}, S.~C.~C. and {Bauer}, S.~M. and {Beaulieu}, M. and {Bellido-Tirado}, O. and {Benneke}, B. and {Bensby}, T. and {Bergin}, E.~A. and {Berio}, P. and {Biazzo}, K. and {Bigot}, L. and {Bik}, A. and {Birkby}, J.~L. and {Blind}, N. and {Boebion}, O. and {Boisse}, I. and {Bolmont}, E. and {Bolton}, J.~S. and {Bonaglia}, M. and {Bonfils}, X. and {Bonhomme}, L. and {Borsa}, F. and {Bouret}, J. -C. and {Brandeker}, A. and {Brandner}, W. and {Broeg}, C.~H. and {Brogi}, M. and {Brousseau}, D. and {Brucalassi}, A. and {Brynnel}, J. and {Buchhave}, L.~A. and {Buscher}, D.~F. and {Cabona}, L. and {Cabral}, A. and {Calderone}, G. and {Calvo-Ortega}, R. and {Cantalloube}, F. and {Canto Martins}, B.~L. and {Carbonaro}, L. and {Caujolle}, Y. and {Chauvin}, G. and {Chazelas}, B. and {Cheffot}, A. -L. and {Cheng}, Y.~S. and {Chiavassa}, A. and {Christensen}, L. and {Cirami}, R. and {Cirasuolo}, M. and {Cook}, N.~J. and {Cooke}, R.~J. and {Coretti}, I. and {Covino}, S. and {Cowan}, N. and {Cresci}, G. and {Cristiani}, S. and {Cunha Parro}, V. and {Cupani}, G. and {D'Odorico}, V. and {Dadi}, K. and {de Castro Le{\~a}o}, I. and {De Cia}, A. and {De Medeiros}, J.~R. and {Debras}, F. and {Debus}, M. and {Delorme}, A. and {Demangeon}, O. and {Derie}, F. and {Dessauges-Zavadsky}, M. and {Di Marcantonio}, P. and {Di Stefano}, S. and {Dionies}, F. and {Domiciano de Souza}, A. and {Doyon}, R. and {Dunn}, J. and {Egner}, S. and {Ehrenreich}, D. and {Faria}, J.~P. and {Ferruzzi}, D. and {Feruglio}, C. and {Fisher}, M. and {Fontana}, A. and {Frank}, B.~S. and {Fuesslein}, C. and {Fumagalli}, M. and {Fusco}, T. and {Fynbo}, J. and {Gabella}, O. and {Gaessler}, W. and {Gallo}, E. and {Gao}, X. and {Genolet}, L. and {Genoni}, M. and {Giacobbe}, P. and {Giro}, E. and {Gon{\c{c}}alves}, R.~S. and {Gonzalez}, O.~A. and {Gonz{\'a}lez-Hern{\'a}ndez}, J.~I. and {Gouvret}, C. and {Gracia T{\'e}mich}, F. and {Haehnelt}, M.~G. and {Haniff}, C. and {Hatzes}, A. and {Helled}, R. and {Hoeijmakers}, H.~J. and {Hughes}, I. and {Huke}, P. and {Ivanisenko}, Y. and {J{\"a}rvinen}, A.~S. and {J{\"a}rvinen}, S.~P. and {Kaminski}, A. and {Kern}, J. and {Knoche}, J. and {Kordt}, A. and {Korhonen}, H. and {Korn}, A.~J. and {Kouach}, D. and {Kowzan}, G. and {Kreidberg}, L. and {Landoni}, M. and {Lanotte}, A.~A. and {Lavail}, A. and {Lavie}, B. and {Lee}, D. and {Lehmitz}, M. and {Li}, J. and {Li}, W. and {Liske}, J. and {Lovis}, C. and {Lucatello}, S. and {Lunney}, D. and {MacIntosh}, M.~J. and {Madhusudhan}, N. and {Magrini}, L. and {Maiolino}, R. and {Maldonado}, J. and {Malo}, L. and {Man}, A.~W.~S. and {Marquart}, T. and {Marques}, C.~M.~J. and {Marques}, E.~L. and {Martinez}, P. and {Martins}, A. and {Martins}, C.~J.~A.~P. and {Martins}, J.~H.~C. and {Maslowski}, P. and {Mason}, C. and {Mason}, E. and {McCracken}, R.~A. and {Melo e Sousa}, M.~A.~F. and {Mergo}, P. and {Micela}, G. and {Milakovi{\'c}}, D. and {Molli{\`e}re}, P. and {Monteiro}, M.~A. and {Montgomery}, D. and {Mordasini}, C. and {Morin}, J. and {Mucciarelli}, A. and {Murphy}, M.~T. and {N'Diaye}, M. and {Nardetto}, N. and {Neichel}, B. and {Neri}, N. and {Niedzielski}, A.~T. and {Niemczura}, E. and {Nisini}, B. and {Nortmann}, L. and {Noterdaeme}, P. and {Nunes}, N.~J. and {Oggioni}, L. and {Olchewsky}, F. and {Oliva}, E. and {{\"O}nel}, H. and {Origlia}, L. and {{\"O}stlin}, G. and {Ouellette}, N.~N. -Q. and {Pall{\'e}}, E. and {Papaderos}, P. and {Pariani}, G. and {Pasquini}, L.},
        title = "{ANDES, the high resolution spectrograph for the ELT: science goals, project overview, and future developments}",
    booktitle = {Ground-based and Airborne Instrumentation for Astronomy X},
         year = 2024,
       editor = {{Bryant}, Julia J. and {Motohara}, Kentaro and {Vernet}, Jo{\"e}l. R.~D.},
       series = {Society of Photo-Optical Instrumentation Engineers (SPIE) Conference Series},
       volume = {13096},
        month = jul,
          eid = {1309613},
        pages = {1309613},
          doi = {10.1117/12.3017966},
archivePrefix = {arXiv},
       eprint = {2407.14601},
 primaryClass = {astro-ph.IM},
       adsurl = {https://ui.adsabs.harvard.edu/abs/2024SPIE13096E..13M}
}

@ARTICLE{ReisPRF:2019AJ....157...16R,
       author = {{Reis}, Itamar and {Baron}, Dalya and {Shahaf}, Sahar},
        title = "{Probabilistic Random Forest: A Machine Learning Algorithm for Noisy Data Sets}",
      journal = {\aj},
         year = 2019,
        month = jan,
       volume = {157},
       number = {1},
          eid = {16},
        pages = {16},
          doi = {10.3847/1538-3881/aaf101},
archivePrefix = {arXiv},
       eprint = {1811.05994},
 primaryClass = {astro-ph.IM},
       adsurl = {https://ui.adsabs.harvard.edu/abs/2019AJ....157...16R}
}

@ARTICLE{CarrascoGaiaSpectra:2021A&A...652A..86C,
       author = {{Carrasco}, J.~M. and {Weiler}, M. and {Jordi}, C. and {Fabricius}, C. and {De Angeli}, F. and {Evans}, D.~W. and {van Leeuwen}, F. and {Riello}, M. and {Montegriffo}, P.},
        title = "{Internal calibration of Gaia BP/RP low-resolution spectra}",
      journal = {\aap},
         year = 2021,
        month = aug,
       volume = {652},
          eid = {A86},
        pages = {A86},
          doi = {10.1051/0004-6361/202141249},
archivePrefix = {arXiv},
       eprint = {2106.01752},
 primaryClass = {astro-ph.IM},
       adsurl = {https://ui.adsabs.harvard.edu/abs/2021A&A...652A..86C}
}

@ARTICLE{GaiaDR3:2023A&A...674A...1G,
       author = {{Gaia Collaboration} and {Vallenari}, A. and {Brown}, A.~G.~A. and {Prusti}, T. and {de Bruijne}, J.~H.~J. and {Arenou}, F. and {Babusiaux}, C. and {Biermann}, M. and {Creevey}, O.~L. and {Ducourant}, C. and {Evans}, D.~W. and {Eyer}, L. and {Guerra}, R. and {Hutton}, A. and {Jordi}, C. and {Klioner}, S.~A. and {Lammers}, U.~L. and {Lindegren}, L. and {Luri}, X. and {Mignard}, F. and {Panem}, C. and {Pourbaix}, D. and {Randich}, S. and {Sartoretti}, P. and {Soubiran}, C. and {Tanga}, P. and {Walton}, N.~A. and {Bailer-Jones}, C.~A.~L. and {Bastian}, U. and {Drimmel}, R. and {Jansen}, F. and {Katz}, D. and {Lattanzi}, M.~G. and {van Leeuwen}, F. and {Bakker}, J. and {Cacciari}, C. and {Casta{\~n}eda}, J. and {De Angeli}, F. and {Fabricius}, C. and {Fouesneau}, M. and {Fr{\'e}mat}, Y. and {Galluccio}, L. and {Guerrier}, A. and {Heiter}, U. and {Masana}, E. and {Messineo}, R. and {Mowlavi}, N. and {Nicolas}, C. and {Nienartowicz}, K. and {Pailler}, F. and {Panuzzo}, P. and {Riclet}, F. and {Roux}, W. and {Seabroke}, G.~M. and {Sordo}, R. and {Th{\'e}venin}, F. and {Gracia-Abril}, G. and {Portell}, J. and {Teyssier}, D. and {Altmann}, M. and {Andrae}, R. and {Audard}, M. and {Bellas-Velidis}, I. and {Benson}, K. and {Berthier}, J. and {Blomme}, R. and {Burgess}, P.~W. and {Busonero}, D. and {Busso}, G. and {C{\'a}novas}, H. and {Carry}, B. and {Cellino}, A. and {Cheek}, N. and {Clementini}, G. and {Damerdji}, Y. and {Davidson}, M. and {de Teodoro}, P. and {Nu{\~n}ez Campos}, M. and {Delchambre}, L. and {Dell'Oro}, A. and {Esquej}, P. and {Fern{\'a}ndez-Hern{\'a}ndez}, J. and {Fraile}, E. and {Garabato}, D. and {Garc{\'\i}a-Lario}, P. and {Gosset}, E. and {Haigron}, R. and {Halbwachs}, J. -L. and {Hambly}, N.~C. and {Harrison}, D.~L. and {Hern{\'a}ndez}, J. and {Hestroffer}, D. and {Hodgkin}, S.~T. and {Holl}, B. and {Jan{\ss}en}, K. and {Jevardat de Fombelle}, G. and {Jordan}, S. and {Krone-Martins}, A. and {Lanzafame}, A.~C. and {L{\"o}ffler}, W. and {Marchal}, O. and {Marrese}, P.~M. and {Moitinho}, A. and {Muinonen}, K. and {Osborne}, P. and {Pancino}, E. and {Pauwels}, T. and {Recio-Blanco}, A. and {Reyl{\'e}}, C. and {Riello}, M. and {Rimoldini}, L. and {Roegiers}, T. and {Rybizki}, J. and {Sarro}, L.~M. and {Siopis}, C. and {Smith}, M. and {Sozzetti}, A. and {Utrilla}, E. and {van Leeuwen}, M. and {Abbas}, U. and {{\'A}brah{\'a}m}, P. and {Abreu Aramburu}, A. and {Aerts}, C. and {Aguado}, J.~J. and {Ajaj}, M. and {Aldea-Montero}, F. and {Altavilla}, G. and {{\'A}lvarez}, M.~A. and {Alves}, J. and {Anders}, F. and {Anderson}, R.~I. and {Anglada Varela}, E. and {Antoja}, T. and {Baines}, D. and {Baker}, S.~G. and {Balaguer-N{\'u}{\~n}ez}, L. and {Balbinot}, E. and {Balog}, Z. and {Barache}, C. and {Barbato}, D. and {Barros}, M. and {Barstow}, M.~A. and {Bartolom{\'e}}, S. and {Bassilana}, J. -L. and {Bauchet}, N. and {Becciani}, U. and {Bellazzini}, M. and {Berihuete}, A. and {Bernet}, M. and {Bertone}, S. and {Bianchi}, L. and {Binnenfeld}, A. and {Blanco-Cuaresma}, S. and {Blazere}, A. and {Boch}, T. and {Bombrun}, A. and {Bossini}, D. and {Bouquillon}, S. and {Bragaglia}, A. and {Bramante}, L. and {Breedt}, E. and {Bressan}, A. and {Brouillet}, N. and {Brugaletta}, E. and {Bucciarelli}, B. and {Burlacu}, A. and {Butkevich}, A.~G. and {Buzzi}, R. and {Caffau}, E. and {Cancelliere}, R. and {Cantat-Gaudin}, T. and {Carballo}, R. and {Carlucci}, T. and {Carnerero}, M.~I. and {Carrasco}, J.~M. and {Casamiquela}, L. and {Castellani}, M. and {Castro-Ginard}, A. and {Chaoul}, L. and {Charlot}, P. and {Chemin}, L. and {Chiaramida}, V. and {Chiavassa}, A. and {Chornay}, N. and {Comoretto}, G. and {Contursi}, G. and {Cooper}, W.~J. and {Cornez}, T. and {Cowell}, S. and {Crifo}, F. and {Cropper}, M. and {Crosta}, M. and {Crowley}, C. and {Dafonte}, C. and {Dapergolas}, A. and {David}, M. and {David}, P. and {de Laverny}, P. and {De Luise}, F. and {De March}, R. and {De Ridder}, J. and {de Souza}, R. and {de Torres}, A. and {del Peloso}, E.~F. and {del Pozo}, E. and {Delbo}, M. and {Delgado}, A. and {Delisle}, J. -B. and {Demouchy}, C. and {Dharmawardena}, T.~E. and {Di Matteo}, P. and {Diakite}, S. and {Diener}, C. and {Distefano}, E. and {Dolding}, C. and {Edvardsson}, B. and {Enke}, H. and {Fabre}, C. and {Fabrizio}, M. and {Faigler}, S. and {Fedorets}, G. and {Fernique}, P. and {Fienga}, A. and {Figueras}, F. and {Fournier}, Y. and {Fouron}, C. and {Fragkoudi}, F. and {Gai}, M. and {Garcia-Gutierrez}, A. and {Garcia-Reinaldos}, M. and {Garc{\'\i}a-Torres}, M. and {Garofalo}, A. and {Gavel}, A. and {Gavras}, P. and {Gerlach}, E. and {Geyer}, R. and {Giacobbe}, P. and {Gilmore}, G. and {Girona}, S. and {Giuffrida}, G. and {Gomel}, R. and {Gomez}, A. and {Gonz{\'a}lez-N{\'u}{\~n}ez}, J. and {Gonz{\'a}lez-Santamar{\'\i}a}, I. and {Gonz{\'a}lez-Vidal}, J.~J. and {Granvik}, M. and {Guillout}, P. and {Guiraud}, J. and {Guti{\'e}rrez-S{\'a}nchez}, R. and {Guy}, L.~P. and {Hatzidimitriou}, D. and {Hauser}, M. and {Haywood}, M. and {Helmer}, A. and {Helmi}, A. and {Sarmiento}, M.~H. and {Hidalgo}, S.~L. and {Hilger}, T. and {H{\l}adczuk}, N. and {Hobbs}, D. and {Holland}, G. and {Huckle}, H.~E. and {Jardine}, K. and {Jasniewicz}, G. and {Jean-Antoine Piccolo}, A. and {Jim{\'e}nez-Arranz}, {\'O}. and {Jorissen}, A. and {Juaristi Campillo}, J. and {Julbe}, F. and {Karbevska}, L. and {Kervella}, P. and {Khanna}, S. and {Kontizas}, M. and {Kordopatis}, G. and {Korn}, A.~J. and {K{\'o}sp{\'a}l}, {\'A}. and {Kostrzewa-Rutkowska}, Z. and {Kruszy{\'n}ska}, K. and {Kun}, M. and {Laizeau}, P. and {Lambert}, S. and {Lanza}, A.~F. and {Lasne}, Y. and {Le Campion}, J. -F. and {Lebreton}, Y. and {Lebzelter}, T. and {Leccia}, S. and {Leclerc}, N. and {Lecoeur-Taibi}, I. and {Liao}, S. and {Licata}, E.~L. and {Lindstr{\o}m}, H.~E.~P. and {Lister}, T.~A. and {Livanou}, E. and {Lobel}, A. and {Lorca}, A. and {Loup}, C. and {Madrero Pardo}, P. and {Magdaleno Romeo}, A. and {Managau}, S. and {Mann}, R.~G. and {Manteiga}, M. and {Marchant}, J.~M. and {Marconi}, M. and {Marcos}, J. and {Marcos Santos}, M.~M.~S. and {Mar{\'\i}n Pina}, D. and {Marinoni}, S. and {Marocco}, F. and {Marshall}, D.~J. and {Martin Polo}, L. and {Mart{\'\i}n-Fleitas}, J.~M. and {Marton}, G. and {Mary}, N. and {Masip}, A. and {Massari}, D. and {Mastrobuono-Battisti}, A. and {Mazeh}, T. and {McMillan}, P.~J. and {Messina}, S. and {Michalik}, D. and {Millar}, N.~R. and {Mints}, A. and {Molina}, D. and {Molinaro}, R. and {Moln{\'a}r}, L. and {Monari}, G. and {Mongui{\'o}}, M. and {Montegriffo}, P. and {Montero}, A. and {Mor}, R. and {Mora}, A. and {Morbidelli}, R. and {Morel}, T. and {Morris}, D. and {Muraveva}, T. and {Murphy}, C.~P. and {Musella}, I. and {Nagy}, Z. and {Noval}, L. and {Oca{\~n}a}, F. and {Ogden}, A. and {Ordenovic}, C. and {Osinde}, J.~O. and {Pagani}, C. and {Pagano}, I. and {Palaversa}, L. and {Palicio}, P.~A. and {Pallas-Quintela}, L. and {Panahi}, A. and {Payne-Wardenaar}, S. and {Pe{\~n}alosa Esteller}, X. and {Penttil{\"a}}, A. and {Pichon}, B. and {Piersimoni}, A.~M. and {Pineau}, F. -X. and {Plachy}, E. and {Plum}, G. and {Poggio}, E. and {Pr{\v{s}}a}, A. and {Pulone}, L. and {Racero}, E. and {Ragaini}, S. and {Rainer}, M. and {Raiteri}, C.~M. and {Rambaux}, N. and {Ramos}, P. and {Ramos-Lerate}, M. and {Re Fiorentin}, P. and {Regibo}, S. and {Richards}, P.~J. and {Rios Diaz}, C. and {Ripepi}, V. and {Riva}, A. and {Rix}, H. -W. and {Rixon}, G. and {Robichon}, N. and {Robin}, A.~C. and {Robin}, C. and {Roelens}, M. and {Rogues}, H.~R.~O. and {Rohrbasser}, L. and {Romero-G{\'o}mez}, M. and {Rowell}, N. and {Royer}, F. and {Ruz Mieres}, D. and {Rybicki}, K.~A. and {Sadowski}, G. and {S{\'a}ez N{\'u}{\~n}ez}, A. and {Sagrist{\`a} Sell{\'e}s}, A. and {Sahlmann}, J. and {Salguero}, E. and {Samaras}, N. and {Sanchez Gimenez}, V. and {Sanna}, N. and {Santove{\~n}a}, R. and {Sarasso}, M. and {Schultheis}, M. and {Sciacca}, E. and {Segol}, M. and {Segovia}, J.~C. and {S{\'e}gransan}, D. and {Semeux}, D. and {Shahaf}, S. and {Siddiqui}, H.~I. and {Siebert}, A. and {Siltala}, L. and {Silvelo}, A. and {Slezak}, E. and {Slezak}, I. and {Smart}, R.~L. and {Snaith}, O.~N. and {Solano}, E. and {Solitro}, F. and {Souami}, D. and {Souchay}, J. and {Spagna}, A. and {Spina}, L. and {Spoto}, F. and {Steele}, I.~A. and {Steidelm{\"u}ller}, H. and {Stephenson}, C.~A. and {S{\"u}veges}, M. and {Surdej}, J. and {Szabados}, L. and {Szegedi-Elek}, E. and {Taris}, F. and {Taylor}, M.~B. and {Teixeira}, R. and {Tolomei}, L. and {Tonello}, N. and {Torra}, F. and {Torra}, J. and {Torralba Elipe}, G. and {Trabucchi}, M. and {Tsounis}, A.~T. and {Turon}, C. and {Ulla}, A. and {Unger}, N. and {Vaillant}, M.~V. and {van Dillen}, E. and {van Reeven}, W. and {Vanel}, O. and {Vecchiato}, A. and {Viala}, Y. and {Vicente}, D. and {Voutsinas}, S. and {Weiler}, M. and {Wevers}, T. and {Wyrzykowski}, {\L}. and {Yoldas}, A. and {Yvard}, P. and {Zhao}, H. and {Zorec}, J. and {Zucker}, S. and {Zwitter}, T.},
        title = "{Gaia Data Release 3. Summary of the content and survey properties}",
      journal = {\aap},
         year = 2023,
        month = jun,
       volume = {674},
          eid = {A1},
        pages = {A1},
          doi = {10.1051/0004-6361/202243940},
archivePrefix = {arXiv},
       eprint = {2208.00211},
 primaryClass = {astro-ph.GA},
       adsurl = {https://ui.adsabs.harvard.edu/abs/2023A&A...674A...1G}
}

@misc{AllWISE_2014yCat.2328....0C,
       author = {{Cutri}, R.~M. and {Wright}, E.~L. and {Conrow}, T. and {Fowler}, J.~W. and {Eisenhardt}, P.~R.~M. and {Grillmair}, C. and {Kirkpatrick}, J.~D. and {Masci}, F. and {McCallon}, H.~L. and {Wheelock}, S.~L. and {Fajardo-Acosta}, S. and {Yan}, L. and {Benford}, D. and {Harbut}, M. and {Jarrett}, T. and {Lake}, S. and {Leisawitz}, D. and {Ressler}, M.~E. and {Stanford}, S.~A. and {Tsai}, C. -W. and {Liu}, F. and {Helou}, G. and {Mainzer}, A. and {Gettngs}, D. and {Gonzalez}, A. and {Hoffman}, D. and {Marsh}, K.~A. and {Padgett}, D. and {Skrutskie}, M.~F. and {Beck}, R. and {Papin}, M. and {Wittman}, M.},
        title = "{VizieR Online Data Catalog: AllWISE Data Release (Cutri+ 2013)}",
 howpublished = {VizieR On-line Data Catalog: II/328.  Originally published in: IPAC/Caltech (2013)},
         year = 2021,
        month = feb,
          eid = {II/328},
       adsurl = {https://ui.adsabs.harvard.edu/abs/2014yCat.2328....0C}
}

@ARTICLE{CatWISE:2021ApJS..253....8M,
       author = {{Marocco}, Federico and {Eisenhardt}, Peter R.~M. and {Fowler}, John W. and {Kirkpatrick}, J. Davy and {Meisner}, Aaron M. and {Schlafly}, Edward F. and {Stanford}, S.~A. and {Garcia}, Nelson and {Caselden}, Dan and {Cushing}, Michael C. and {Cutri}, Roc M. and {Faherty}, Jacqueline K. and {Gelino}, Christopher R. and {Gonzalez}, Anthony H. and {Jarrett}, Thomas H. and {Koontz}, Renata and {Mainzer}, Amanda and {Marchese}, Elijah J. and {Mobasher}, Bahram and {Schlegel}, David J. and {Stern}, Daniel and {Teplitz}, Harry I. and {Wright}, Edward L.},
        title = "{The CatWISE2020 Catalog}",
      journal = {\apjs},
         year = 2021,
        month = mar,
       volume = {253},
       number = {1},
          eid = {8},
        pages = {8},
          doi = {10.3847/1538-4365/abd805},
archivePrefix = {arXiv},
       eprint = {2012.13084},
 primaryClass = {astro-ph.IM},
       adsurl = {https://ui.adsabs.harvard.edu/abs/2021ApJS..253....8M}
}

@ARTICLE{WISE:2010AJ....140.1868W,
       author = {{Wright}, Edward L. and {Eisenhardt}, Peter R.~M. and {Mainzer}, Amy K. and {Ressler}, Michael E. and {Cutri}, Roc M. and {Jarrett}, Thomas and {Kirkpatrick}, J. Davy and {Padgett}, Deborah and {McMillan}, Robert S. and {Skrutskie}, Michael and {Stanford}, S.~A. and {Cohen}, Martin and {Walker}, Russell G. and {Mather}, John C. and {Leisawitz}, David and {Gautier}, Thomas N., III and {McLean}, Ian and {Benford}, Dominic and {Lonsdale}, Carol J. and {Blain}, Andrew and {Mendez}, Bryan and {Irace}, William R. and {Duval}, Valerie and {Liu}, Fengchuan and {Royer}, Don and {Heinrichsen}, Ingolf and {Howard}, Joan and {Shannon}, Mark and {Kendall}, Martha and {Walsh}, Amy L. and {Larsen}, Mark and {Cardon}, Joel G. and {Schick}, Scott and {Schwalm}, Mark and {Abid}, Mohamed and {Fabinsky}, Beth and {Naes}, Larry and {Tsai}, Chao-Wei},
        title = "{The Wide-field Infrared Survey Explorer (WISE): Mission Description and Initial On-orbit Performance}",
      journal = {\aj},
         year = 2010,
        month = dec,
       volume = {140},
       number = {6},
        pages = {1868-1881},
          doi = {10.1088/0004-6256/140/6/1868},
archivePrefix = {arXiv},
       eprint = {1008.0031},
 primaryClass = {astro-ph.IM},
       adsurl = {https://ui.adsabs.harvard.edu/abs/2010AJ....140.1868W}
}

@ARTICLE{NEOWISE:2014ApJ...792...30M,
       author = {{Mainzer}, A. and {Bauer}, J. and {Cutri}, R.~M. and {Grav}, T. and {Masiero}, J. and {Beck}, R. and {Clarkson}, P. and {Conrow}, T. and {Dailey}, J. and {Eisenhardt}, P. and {Fabinsky}, B. and {Fajardo-Acosta}, S. and {Fowler}, J. and {Gelino}, C. and {Grillmair}, C. and {Heinrichsen}, I. and {Kendall}, M. and {Kirkpatrick}, J. Davy and {Liu}, F. and {Masci}, F. and {McCallon}, H. and {Nugent}, C.~R. and {Papin}, M. and {Rice}, E. and {Royer}, D. and {Ryan}, T. and {Sevilla}, P. and {Sonnett}, S. and {Stevenson}, R. and {Thompson}, D.~B. and {Wheelock}, S. and {Wiemer}, D. and {Wittman}, M. and {Wright}, E. and {Yan}, L.},
        title = "{Initial Performance of the NEOWISE Reactivation Mission}",
      journal = {\apj},
         year = 2014,
        month = sep,
       volume = {792},
       number = {1},
          eid = {30},
        pages = {30},
          doi = {10.1088/0004-637X/792/1/30},
archivePrefix = {arXiv},
       eprint = {1406.6025},
 primaryClass = {astro-ph.EP},
       adsurl = {https://ui.adsabs.harvard.edu/abs/2014ApJ...792...30M}
}

@ARTICLE{Veron10:2010A&A...518A..10V,
       author = {{V{\'e}ron-Cetty}, M. -P. and {V{\'e}ron}, P.},
        title = "{A catalogue of quasars and active nuclei: 13th edition}",
      journal = {\aap},
         year = 2010,
        month = jul,
       volume = {518},
          eid = {A10},
        pages = {A10},
          doi = {10.1051/0004-6361/201014188},
       adsurl = {https://ui.adsabs.harvard.edu/abs/2010A&A...518A..10V}
}

@ARTICLE{2df:2001MNRAS.328.1039C,
       author = {{Colless}, Matthew and {Dalton}, Gavin and {Maddox}, Steve and {Sutherland}, Will and {Norberg}, Peder and {Cole}, Shaun and {Bland-Hawthorn}, Joss and {Bridges}, Terry and {Cannon}, Russell and {Collins}, Chris and {Couch}, Warrick and {Cross}, Nicholas and {Deeley}, Kathryn and {De Propris}, Roberto and {Driver}, Simon P. and {Efstathiou}, George and {Ellis}, Richard S. and {Frenk}, Carlos S. and {Glazebrook}, Karl and {Jackson}, Carole and {Lahav}, Ofer and {Lewis}, Ian and {Lumsden}, Stuart and {Madgwick}, Darren and {Peacock}, John A. and {Peterson}, Bruce A. and {Price}, Ian and {Seaborne}, Mark and {Taylor}, Keith},
        title = "{The 2dF Galaxy Redshift Survey: spectra and redshifts}",
      journal = {\mnras},
         year = 2001,
        month = dec,
       volume = {328},
       number = {4},
        pages = {1039-1063},
          doi = {10.1046/j.1365-8711.2001.04902.x},
archivePrefix = {arXiv},
       eprint = {astro-ph/0106498},
 primaryClass = {astro-ph},
       adsurl = {https://ui.adsabs.harvard.edu/abs/2001MNRAS.328.1039C}
}

@ARTICLE{Liske+08:2008MNRAS.386.1192L,
       author = {{Liske}, J. and {Grazian}, A. and {Vanzella}, E. and {Dessauges}, M. and {Viel}, M. and {Pasquini}, L. and {Haehnelt}, M. and {Cristiani}, S. and {Pepe}, F. and {Avila}, G. and {Bonifacio}, P. and {Bouchy}, F. and {Dekker}, H. and {Delabre}, B. and {D'Odorico}, S. and {D'Odorico}, V. and {Levshakov}, S. and {Lovis}, C. and {Mayor}, M. and {Molaro}, P. and {Moscardini}, L. and {Murphy}, M.~T. and {Queloz}, D. and {Shaver}, P. and {Udry}, S. and {Wiklind}, T. and {Zucker}, S.},
        title = "{Cosmic dynamics in the era of Extremely Large Telescopes}",
      journal = {\mnras},
         year = 2008,
        month = may,
       volume = {386},
       number = {3},
        pages = {1192-1218},
          doi = {10.1111/j.1365-2966.2008.13090.x},
archivePrefix = {arXiv},
       eprint = {0802.1532},
 primaryClass = {astro-ph},
       adsurl = {https://ui.adsabs.harvard.edu/abs/2008MNRAS.386.1192L}
}

@manual{LSST_guy_2025_15558559,
  title        = {Rubin Observatory Plans for an Early Science
                   Program
                  },
  author       = {Guy, Leanne  P. and
                  Bechtol, Keith and
                  Bellm, Eric and
                  Blum, Bob and
                  Graham, Melissa L. and
                  Ivezić, Željko and
                  Lupton, Robert and
                  Marshall, Phil and
                  Slater, Colin T. and
                  Strauss, Michael and
                  Dubois-Felsmann, Gregory},
  month        = may,
  year         = 2025,
  doi          = {10.5281/zenodo.15558559},
  url          = {https://doi.org/10.5281/zenodo.15558559},
}

@ARTICLE{Euclid_2025A&A...697A...1E,
       author = {{Euclid Collaboration} and {Mellier}, Y. and {Abdurro'uf} and {Acevedo Barroso}, J.~A. and {Ach{\'u}carro}, A. and {Adamek}, J. and {Adam}, R. and {Addison}, G.~E. and {Aghanim}, N. and {Aguena}, M. and et al.},
        title = "{Euclid: I. Overview of the Euclid mission}",
      journal = {\aap},
         year = 2025,
        month = may,
       volume = {697},
          eid = {A1},
        pages = {A1},
          doi = {10.1051/0004-6361/202450810},
archivePrefix = {arXiv},
       eprint = {2405.13491},
 primaryClass = {astro-ph.CO},
       adsurl = {https://ui.adsabs.harvard.edu/abs/2025A&A...697A...1E}
}

@ARTICLE{Euclid:2022A&A...662A.112E,
       author = {{Euclid Collaboration} and {Scaramella}, R. and {Amiaux}, J. and {Mellier}, Y. and {Burigana}, C. and {Carvalho}, C.~S. and {Cuillandre}, J.-C. and {Da Silva}, A. and {Derosa}, A. and {Dinis}, J. and {Maiorano}, E. and {Maris}, M. and {Tereno}, I. and {Laureijs}, R. and {Boenke}, T. and {Buenadicha}, G. and {Dupac}, X. and {Gaspar Venancio}, L.~M. and {G{\'o}mez-{\'A}lvarez}, P. and {Hoar}, J. and {Lorenzo Alvarez}, J. and {Racca}, G.~D. and {Saavedra-Criado}, G. and {Schwartz}, J. and {Vavrek}, R. and {Schirmer}, M. and {Aussel}, H. and {Azzollini}, R. and {Cardone}, V.~F. and {Cropper}, M. and {Ealet}, A. and {Garilli}, B. and {Gillard}, W. and {Granett}, B.~R. and {Guzzo}, L. and {Hoekstra}, H. and {Jahnke}, K. and {Kitching}, T. and {Maciaszek}, T. and {Meneghetti}, M. and {Miller}, L. and {Nakajima}, R. and {Niemi}, S.~M. and {Pasian}, F. and {Percival}, W.~J. and {Pottinger}, S. and {Sauvage}, M. and {Scodeggio}, M. and {Wachter}, S. and {Zacchei}, A. and {Aghanim}, N. and {Amara}, A. and {Auphan}, T. and {Auricchio}, N. and {Awan}, S. and {Balestra}, A. and {Bender}, R. and {Bodendorf}, C. and {Bonino}, D. and {Branchini}, E. and {Brau-Nogue}, S. and {Brescia}, M. and {Candini}, G.~P. and {Capobianco}, V. and {Carbone}, C. and {Carlberg}, R.~G. and {Carretero}, J. and {Casas}, R. and {Castander}, F.~J. and {Castellano}, M. and {Cavuoti}, S. and {Cimatti}, A. and {Cledassou}, R. and {Congedo}, G. and {Conselice}, C.~J. and {Conversi}, L. and {Copin}, Y. and {Corcione}, L. and {Costille}, A. and {Courbin}, F. and {Degaudenzi}, H. and {Douspis}, M. and {Dubath}, F. and {Duncan}, C.~A.~J. and {Dusini}, S. and {Farrens}, S. and {Ferriol}, S. and {Fosalba}, P. and {Fourmanoit}, N. and {Frailis}, M. and {Franceschi}, E. and {Franzetti}, P. and {Fumana}, M. and {Gillis}, B. and {Giocoli}, C. and {Grazian}, A. and {Grupp}, F. and {Haugan}, S.~V.~H. and {Holmes}, W. and {Hormuth}, F. and {Hudelot}, P. and {Kermiche}, S. and {Kiessling}, A. and {Kilbinger}, M. and {Kohley}, R. and {Kubik}, B. and {K{\"u}mmel}, M. and {Kunz}, M. and {Kurki-Suonio}, H. and {Lahav}, O. and {Ligori}, S. and {Lilje}, P.~B. and {Lloro}, I. and {Mansutti}, O. and {Marggraf}, O. and {Markovic}, K. and {Marulli}, F. and {Massey}, R. and {Maurogordato}, S. and {Melchior}, M. and {Merlin}, E. and {Meylan}, G. and {Mohr}, J.~J. and {Moresco}, M. and {Morin}, B. and {Moscardini}, L. and {Munari}, E. and {Nichol}, R.~C. and {Padilla}, C. and {Paltani}, S. and {Peacock}, J. and {Pedersen}, K. and {Pettorino}, V. and {Pires}, S. and {Poncet}, M. and {Popa}, L. and {Pozzetti}, L. and {Raison}, F. and {Rebolo}, R. and {Rhodes}, J. and {Rix}, H.-W. and {Roncarelli}, M. and {Rossetti}, E. and {Saglia}, R. and {Schneider}, P. and {Schrabback}, T. and {Secroun}, A. and {Seidel}, G. and {Serrano}, S. and {Sirignano}, C. and {Sirri}, G. and {Skottfelt}, J. and {Stanco}, L. and {Starck}, J.~L. and {Tallada-Cresp{\'\i}}, P. and {Tavagnacco}, D. and {Taylor}, A.~N. and {Teplitz}, H.~I. and {Toledo-Moreo}, R. and {Torradeflot}, F. and {Trifoglio}, M. and {Valentijn}, E.~A. and {Valenziano}, L. and {Verdoes Kleijn}, G.~A. and {Wang}, Y. and {Welikala}, N. and {Weller}, J. and {Wetzstein}, M. and {Zamorani}, G. and {Zoubian}, J. and {Andreon}, S. and {Baldi}, M. and {Bardelli}, S. and {Boucaud}, A. and {Camera}, S. and {Di Ferdinando}, D. and {Fabbian}, G. and {Farinelli}, R. and {Galeotta}, S. and {Graci{\'a}-Carpio}, J. and {Maino}, D. and {Medinaceli}, E. and {Mei}, S. and {Neissner}, C. and {Polenta}, G. and {Renzi}, A. and {Romelli}, E. and {Rosset}, C. and {Sureau}, F. and {Tenti}, M. and {Vassallo}, T. and {Zucca}, E. and {Baccigalupi}, C. and {Balaguera-Antol{\'\i}nez}, A. and {Battaglia}, P. and {Biviano}, A. and {Borgani}, S. and {Bozzo}, E. and {Cabanac}, R. and {Cappi}, A.},
        title = "{Euclid preparation. I. The Euclid Wide Survey}",
      journal = {\aap},
         year = 2022,
        month = jun,
       volume = {662},
          eid = {A112},
        pages = {A112},
          doi = {10.1051/0004-6361/202141938},
archivePrefix = {arXiv},
       eprint = {2108.01201},
 primaryClass = {astro-ph.CO},
       adsurl = {https://ui.adsabs.harvard.edu/abs/2022A&A...662A.112E}
}

@ARTICLE{RomanSpaceTelecope:2022ApJ...928....1W,
       author = {{Wang}, Yun and {Zhai}, Zhongxu and {Alavi}, Anahita and {Massara}, Elena and {Pisani}, Alice and {Benson}, Andrew and {Hirata}, Christopher M. and {Samushia}, Lado and {Weinberg}, David H. and {Colbert}, James and {Dor{\'e}}, Olivier and {Eifler}, Tim and {Heinrich}, Chen and {Ho}, Shirley and {Krause}, Elisabeth and {Padmanabhan}, Nikhil and {Spergel}, David and {Teplitz}, Harry I.},
        title = "{The High Latitude Spectroscopic Survey on the Nancy Grace Roman Space Telescope}",
      journal = {\apj},
         year = 2022,
        month = mar,
       volume = {928},
       number = {1},
          eid = {1},
        pages = {1},
          doi = {10.3847/1538-4357/ac4973},
archivePrefix = {arXiv},
       eprint = {2110.01829},
 primaryClass = {astro-ph.CO},
       adsurl = {https://ui.adsabs.harvard.edu/abs/2022ApJ...928....1W}
}

@ARTICLE{Wolf20QSO:2020MNRAS.491.1970W,
       author = {{Wolf}, Christian and {Hon}, Wei Jeat and {Bian}, Fuyan and {Onken}, Christopher A. and {Alonzi}, Noura and {Bessell}, Michael A. and {Li}, Zefeng and {Schmidt}, Brian P. and {Tisserand}, Patrick},
        title = "{Ultra-luminous quasars at redshift z > 4.5 from SkyMapper}",
      journal = {\mnras},
         year = 2020,
        month = jan,
       volume = {491},
       number = {2},
        pages = {1970-1979},
          doi = {10.1093/mnras/stz2955},
archivePrefix = {arXiv},
       eprint = {1910.05856},
 primaryClass = {astro-ph.GA},
       adsurl = {https://ui.adsabs.harvard.edu/abs/2020MNRAS.491.1970W}
}

@ARTICLE{Onken2022:2022MNRAS.511..572O,
       author = {{Onken}, Christopher A. and {Wolf}, Christian and {Bian}, Fuyan and {Fan}, Xiaohui and {Hon}, Wei Jeat and {Raithel}, David and {Tisserand}, Patrick and {Lai}, Samuel},
        title = "{Ultraluminous high-redshift quasars from SkyMapper - II. New quasars and the bright end of the luminosity function}",
      journal = {\mnras},
         year = 2022,
        month = mar,
       volume = {511},
       number = {1},
        pages = {572-594},
          doi = {10.1093/mnras/stac051},
archivePrefix = {arXiv},
       eprint = {2105.12215},
 primaryClass = {astro-ph.CO},
       adsurl = {https://ui.adsabs.harvard.edu/abs/2022MNRAS.511..572O}
}

@ARTICLE{Yang2023:2023ApJS..269...27Y,
       author = {{Yang}, Jinyi and {Fan}, Xiaohui and {Gupta}, Ansh and {Myers}, Adam D. and {Palanque-Delabrouille}, Nathalie and {Wang}, Feige and {Y{\`e}che}, Christophe and {Aguilar}, Jessica Nicole and {Ahlen}, Steven and {Alexander}, David M. and {Brooks}, David and {Dawson}, Kyle and {de la Macorra}, Axel and {Dey}, Arjun and {Dhungana}, Govinda and {Fanning}, Kevin and {Font-Ribera}, Andreu and {Gontcho}, Satya and {Guy}, Julien and {Honscheid}, Klaus and {Juneau}, Stephanie and {Kisner}, Theodore and {Kremin}, Anthony and {Le Guillou}, Laurent and {Levi}, Michael and {Magneville}, Christophe and {Martini}, Paul and {Meisner}, Aaron and {Miquel}, Ramon and {Moustakas}, John and {Nie}, Jundan and {Percival}, Will and {Poppett}, Claire and {Prada}, Francisco and {Schlafly}, Edward and {Tarl{\'e}}, Gregory and {Vargas Magana}, Mariana and {Weaver}, Benjamin Alan and {Wechsler}, Risa and {Zhou}, Rongpu and {Zhou}, Zhimin and {Zou}, Hu},
        title = "{DESI z {\ensuremath{\gtrsim}} 5 Quasar Survey. I. A First Sample of 400 New Quasars at z 4.7-6.6}",
      journal = {\apjs},
         year = 2023,
        month = nov,
       volume = {269},
       number = {1},
          eid = {27},
        pages = {27},
          doi = {10.3847/1538-4365/acf99b},
archivePrefix = {arXiv},
       eprint = {2302.01777},
 primaryClass = {astro-ph.GA},
       adsurl = {https://ui.adsabs.harvard.edu/abs/2023ApJS..269...27Y}
}

@ARTICLE{DESI-EDR1:2024AJ....168...58D,
       author = {{DESI Collaboration} and {Adame}, A.~G. and {Aguilar}, J. and {Ahlen}, S. and {Alam}, S. and {Aldering}, G. and {Alexander}, D.~M. and {Alfarsy}, R. and {Allende Prieto}, C. and {Alvarez}, M. and {Alves}, O. and {Anand}, A. and {Andrade-Oliveira}, F. and {Armengaud}, E. and {Asorey}, J. and {Avila}, S. and {Aviles}, A. and {Bailey}, S. and {Balaguera-Antol{\'\i}nez}, A. and {Ballester}, O. and {Baltay}, C. and {Bault}, A. and {Bautista}, J. and {Behera}, J. and {Beltran}, S.~F. and {BenZvi}, S. and {Beraldo e Silva}, L. and {Bermejo-Climent}, J.~R. and {Berti}, A. and {Besuner}, R. and {Beutler}, F. and {Bianchi}, D. and {Blake}, C. and {Blum}, R. and {Bolton}, A.~S. and {Brieden}, S. and {Brodzeller}, A. and {Brooks}, D. and {Brown}, Z. and {Buckley-Geer}, E. and {Burtin}, E. and {Cabayol-Garcia}, L. and {Cai}, Z. and {Canning}, R. and {Cardiel-Sas}, L. and {Carnero Rosell}, A. and {Castander}, F.~J. and {Cervantes-Cota}, J.~L. and {Chabanier}, S. and {Chaussidon}, E. and {Chaves-Montero}, J. and {Chen}, S. and {Chen}, X. and {Chuang}, C. and {Claybaugh}, T. and {Cole}, S. and {Cooper}, A.~P. and {Cuceu}, A. and {Davis}, T.~M. and {Dawson}, K. and {de Belsunce}, R. and {de la Cruz}, R. and {de la Macorra}, A. and {Della Costa}, J. and {de Mattia}, A. and {Demina}, R. and {Demirbozan}, U. and {DeRose}, J. and {Dey}, A. and {Dey}, B. and {Dhungana}, G. and {Ding}, J. and {Ding}, Z. and {Doel}, P. and {Doshi}, R. and {Douglass}, K. and {Edge}, A. and {Eftekharzadeh}, S. and {Eisenstein}, D.~J. and {Elliott}, A. and {Ereza}, J. and {Escoffier}, S. and {Fagrelius}, P. and {Fan}, X. and {Fanning}, K. and {Fawcett}, V.~A. and {Ferraro}, S. and {Flaugher}, B. and {Font-Ribera}, A. and {Forero-Romero}, J.~E. and {Forero-S{\'a}nchez}, D. and {Frenk}, C.~S. and {G{\"a}nsicke}, B.~T. and {Garc{\'\i}a}, L. {\'A}. and {Garc{\'\i}a-Bellido}, J. and {Garcia-Quintero}, C. and {Garrison}, L.~H. and {Gil-Mar{\'\i}n}, H. and {Golden-Marx}, J. and {Gontcho A Gontcho}, S. and {Gonzalez-Morales}, A.~X. and {Gonzalez-Perez}, V. and {Gordon}, C. and {Graur}, O. and {Green}, D. and {Gruen}, D. and {Guy}, J. and {Hadzhiyska}, B. and {Hahn}, C. and {Han}, J.~J. and {Hanif}, M.~M.~S. and {Herrera-Alcantar}, H.~K. and {Honscheid}, K. and {Hou}, J. and {Howlett}, C. and {Huterer}, D. and {Ir{\v{s}}i{\v{c}}}, V. and {Ishak}, M. and {Jacques}, A. and {Jana}, A. and {Jiang}, L. and {Jimenez}, J. and {Jing}, Y.~P. and {Joudaki}, S. and {Joyce}, R. and {Jullo}, E. and {Juneau}, S. and {Kara{\c{c}}ayl{\i}}, N.~G. and {Karim}, T. and {Kehoe}, R. and {Kent}, S. and {Khederlarian}, A. and {Kim}, S. and {Kirkby}, D. and {Kisner}, T. and {Kitaura}, F. and {Kizhuprakkat}, N. and {Kneib}, J. and {Koposov}, S.~E. and {Kov{\'a}cs}, A. and {Kremin}, A. and {Krolewski}, A. and {L'Huillier}, B. and {Lahav}, O. and {Lambert}, A. and {Lamman}, C. and {Lan}, T.-W. and {Landriau}, M. and {Lang}, D. and {Lange}, J.~U. and {Lasker}, J. and {Leauthaud}, A. and {Le Guillou}, L. and {Levi}, M.~E. and {Li}, T.~S. and {Linder}, E. and {Lyons}, A. and {Magneville}, C. and {Manera}, M. and {Manser}, C.~J. and {Margala}, D. and {Martini}, P. and {McDonald}, P. and {Medina}, G.~E. and {Medina-Varela}, L. and {Meisner}, A. and {Mena-Fern{\'a}ndez}, J. and {Meneses-Rizo}, J. and {Mezcua}, M. and {Miquel}, R. and {Montero-Camacho}, P. and {Moon}, J. and {Moore}, S. and {Moustakas}, J. and {Mueller}, E. and {Mundet}, J. and {Mu{\~n}oz-Guti{\'e}rrez}, A. and {Myers}, A.~D. and {Nadathur}, S. and {Napolitano}, L. and {Neveux}, R. and {Newman}, J.~A. and {Nie}, J. and {Nikutta}, R. and {Niz}, G. and {Norberg}, P. and {Noriega}, H.~E. and {Paillas}, E. and {Palanque-Delabrouille}, N. and {Palmese}, A. and {Pan}, Z. and {Parkinson}, D. and {Penmetsa}, S. and {Percival}, W.~J. and {P{\'e}rez-Fern{\'a}ndez}, A. and {P{\'e}rez-R{\`a}fols}, I. and {Pieri}, M. and {Poppett}, C. and {Porredon}, A. and {Pothier}, S.},
        title = "{The Early Data Release of the Dark Energy Spectroscopic Instrument}",
      journal = {\aj},
         year = 2024,
        month = aug,
       volume = {168},
       number = {2},
          eid = {58},
        pages = {58},
          doi = {10.3847/1538-3881/ad3217},
archivePrefix = {arXiv},
       eprint = {2306.06308},
 primaryClass = {astro-ph.CO},
       adsurl = {https://ui.adsabs.harvard.edu/abs/2024AJ....168...58D}
}

@ARTICLE{LAMOST-DR6-9:2023ApJS..265...25J,
       author = {{Jin}, Jun-Jie and {Wu}, Xue-Bing and {Fu}, Yuming and {Yao}, Su and {Ai}, Yan-Li and {Feng}, Xiao-Tong and {He}, Zi-Qi and {Ma}, Qin-Chun and {Pang}, Yu-Xuan and {Zhu}, Rui and {Zhang}, Yan-xia and {Yuan}, Hai-long and {Huo}, Zhi-Ying},
        title = "{The Large Sky Area Multi-Object Fiber Spectroscopic Telescope (LAMOST) Quasar Survey: Quasar Properties from Data Releases 6 to 9}",
      journal = {\apjs},
         year = 2023,
        month = mar,
       volume = {265},
       number = {1},
          eid = {25},
        pages = {25},
          doi = {10.3847/1538-4365/acaf89},
archivePrefix = {arXiv},
       eprint = {2212.12876},
 primaryClass = {astro-ph.GA},
       adsurl = {https://ui.adsabs.harvard.edu/abs/2023ApJS..265...25J}
}

@ARTICLE{LAMOST-DR4-5:2019ApJS..240....6Y,
       author = {{Yao}, Su and {Wu}, Xue-Bing and {Ai}, Y.~L. and {Yang}, Jinyi and {Yang}, Qian and {Dong}, Xiaoyi and {Joshi}, Ravi and {Wang}, Feige and {Feng}, Xiaotong and {Fu}, Yuming and {Hou}, Wen and {Luo}, A.-L. and {Kong}, Xiao and {Liu}, Yuanqi and {Zhao}, Y.-H. and {Zhang}, Y.-X. and {Yuan}, H.-L. and {Shen}, Shiyin},
        title = "{The Large Sky Area Multi-object Fiber Spectroscopic Telescope (LAMOST) Quasar Survey: The Fourth and Fifth Data Releases}",
      journal = {\apjs},
         year = 2019,
        month = jan,
       volume = {240},
       number = {1},
          eid = {6},
        pages = {6},
          doi = {10.3847/1538-4365/aaef88},
archivePrefix = {arXiv},
       eprint = {1811.01570},
 primaryClass = {astro-ph.GA},
       adsurl = {https://ui.adsabs.harvard.edu/abs/2019ApJS..240....6Y}
}

@ARTICLE{SDSS-PhotZ:2009ApJS..180...67R,
       author = {{Richards}, Gordon T. and {Myers}, Adam D. and {Gray}, Alexander G. and {Riegel}, Ryan N. and {Nichol}, Robert C. and {Brunner}, Robert J. and {Szalay}, Alexander S. and {Schneider}, Donald P. and {Anderson}, Scott F.},
        title = "{Efficient Photometric Selection of Quasars from the Sloan Digital Sky Survey. II. \raisebox{-0.5ex}\textasciitilde1,000,000 Quasars from Data Release 6}",
      journal = {\apjs},
         year = 2009,
        month = jan,
       volume = {180},
       number = {1},
        pages = {67-83},
          doi = {10.1088/0067-0049/180/1/67},
archivePrefix = {arXiv},
       eprint = {0809.3952},
 primaryClass = {astro-ph},
       adsurl = {https://ui.adsabs.harvard.edu/abs/2009ApJS..180...67R}
}

@ARTICLE{DESDR2:2021ApJS..255...20A,
       author = {{Abbott}, T.~M.~C. and {Adam{\'o}w}, M. and {Aguena}, M. and {Allam}, S. and {Amon}, A. and {Annis}, J. and {Avila}, S. and {Bacon}, D. and {Banerji}, M. and {Bechtol}, K. and {Becker}, M.~R. and {Bernstein}, G.~M. and {Bertin}, E. and {Bhargava}, S. and {Bridle}, S.~L. and {Brooks}, D. and {Burke}, D.~L. and {Carnero Rosell}, A. and {Carrasco Kind}, M. and {Carretero}, J. and {Castander}, F.~J. and {Cawthon}, R. and {Chang}, C. and {Choi}, A. and {Conselice}, C. and {Costanzi}, M. and {Crocce}, M. and {da Costa}, L.~N. and {Davis}, T.~M. and {De Vicente}, J. and {DeRose}, J. and {Desai}, S. and {Diehl}, H.~T. and {Dietrich}, J.~P. and {Drlica-Wagner}, A. and {Eckert}, K. and {Elvin-Poole}, J. and {Everett}, S. and {Evrard}, A.~E. and {Ferrero}, I. and {Fert{\'e}}, A. and {Flaugher}, B. and {Fosalba}, P. and {Friedel}, D. and {Frieman}, J. and {Garc{\'\i}a-Bellido}, J. and {Gaztanaga}, E. and {Gelman}, L. and {Gerdes}, D.~W. and {Giannantonio}, T. and {Gill}, M.~S.~S. and {Gruen}, D. and {Gruendl}, R.~A. and {Gschwend}, J. and {Gutierrez}, G. and {Hartley}, W.~G. and {Hinton}, S.~R. and {Hollowood}, D.~L. and {Honscheid}, K. and {Huterer}, D. and {James}, D.~J. and {Jeltema}, T. and {Johnson}, M.~D. and {Kent}, S. and {Kron}, R. and {Kuehn}, K. and {Kuropatkin}, N. and {Lahav}, O. and {Li}, T.~S. and {Lidman}, C. and {Lin}, H. and {MacCrann}, N. and {Maia}, M.~A.~G. and {Manning}, T.~A. and {Maloney}, J.~D. and {March}, M. and {Marshall}, J.~L. and {Martini}, P. and {Melchior}, P. and {Menanteau}, F. and {Miquel}, R. and {Morgan}, R. and {Myles}, J. and {Neilsen}, E. and {Ogando}, R.~L.~C. and {Palmese}, A. and {Paz-Chinch{\'o}n}, F. and {Petravick}, D. and {Pieres}, A. and {Plazas}, A.~A. and {Pond}, C. and {Rodriguez-Monroy}, M. and {Romer}, A.~K. and {Roodman}, A. and {Rykoff}, E.~S. and {Sako}, M. and {Sanchez}, E. and {Santiago}, B. and {Scarpine}, V. and {Serrano}, S. and {Sevilla-Noarbe}, I. and {Smith}, J. Allyn and {Smith}, M. and {Soares-Santos}, M. and {Suchyta}, E. and {Swanson}, M.~E.~C. and {Tarle}, G. and {Thomas}, D. and {To}, C. and {Tremblay}, P.~E. and {Troxel}, M.~A. and {Tucker}, D.~L. and {Turner}, D.~J. and {Varga}, T.~N. and {Walker}, A.~R. and {Wechsler}, R.~H. and {Weller}, J. and {Wester}, W. and {Wilkinson}, R.~D. and {Yanny}, B. and {Zhang}, Y. and {Nikutta}, R. and {Fitzpatrick}, M. and {Jacques}, A. and {Scott}, A. and {Olsen}, K. and {Huang}, L. and {Herrera}, D. and {Juneau}, S. and {Nidever}, D. and {Weaver}, B.~A. and {Adean}, C. and {Correia}, V. and {de Freitas}, M. and {Freitas}, F.~N. and {Singulani}, C. and {Vila-Verde}, G. and {Linea Science Server}},
        title = "{The Dark Energy Survey Data Release 2}",
      journal = {\apjs},
         year = 2021,
        month = aug,
       volume = {255},
       number = {2},
          eid = {20},
        pages = {20},
          doi = {10.3847/1538-4365/ac00b3},
archivePrefix = {arXiv},
       eprint = {2101.05765},
 primaryClass = {astro-ph.IM},
       adsurl = {https://ui.adsabs.harvard.edu/abs/2021ApJS..255...20A}
}

@ARTICLE{ref:cupaniFeLoBALs,
       author = {{Cupani}, Guido and {Calderone}, Giorgio and {Selvelli}, Pierluigi and {Cristiani}, Stefano and {Boutsia}, Konstantina and {Grazian}, Andrea and {Fontanot}, Fabio and {Guarneri}, Francesco and {D'Odorico}, Valentina and {Giallongo}, Emanuele and {Menci}, Nicola},
        title = "{Near-infrared spectroscopy of extreme BAL QSOs from the QUBRICS bright quasar survey}",
      journal = {\mnras},
         year = 2022,
        month = feb,
       volume = {510},
       number = {2},
        pages = {2509-2528},
          doi = {10.1093/mnras/stab3562},
archivePrefix = {arXiv},
       eprint = {2112.02594},
 primaryClass = {astro-ph.CO},
       adsurl = {https://ui.adsabs.harvard.edu/abs/2022MNRAS.510.2509C}
}

@book{ref:CCA,
	title =     {An Introduction to Multivariate Statistical Analysis},
	author =    {Theodore W. Anderson},
	publisher = {Wiley},
	isbn =      {978-0-471-36091-9},
	year =      {2003},
	series =    {Wiley series in probability and mathematical statistics},
	edition =   {3},
	volume =    {},
	url =       {https://www.wiley.com/en-us/An+Introduction+to+Multivariate+Statistical+Analysis%2C+3rd+Edition-p-9780471360919}}

@INPROCEEDINGS{Trakhtenbrot2021,
       author = {{Trakhtenbrot}, Benny},
        title = "{What do observations tell us about the highest-redshift supermassive black holes?}",
    booktitle = {Nuclear Activity in Galaxies Across Cosmic Time},
         year = 2021,
       editor = {{Povi{\'c}}, Mirjana and {Marziani}, Paola and {Masegosa}, Josefa and {Netzer}, Hagai and {Negu}, Seblu H. and {Tessema}, Solomon B.},
       volume = {356},
        month = jan,
        pages = {261-275},
          doi = {10.1017/S1743921320003087},
archivePrefix = {arXiv},
       eprint = {2002.00972},
 primaryClass = {astro-ph.GA},
       adsurl = {https://ui.adsabs.harvard.edu/abs/2021IAUS..356..261T}
}

@ARTICLE{Fan2023,
       author = {{Fan}, Liuyuan and {Fang}, Guanwen and {Hu}, Jian},
        title = "{A comparison of cosmological models with high-redshift quasars}",
      journal = {\apss},
         year = 2023,
        month = jul,
       volume = {368},
       number = {7},
          eid = {59},
        pages = {59},
          doi = {10.1007/s10509-023-04215-0},
archivePrefix = {arXiv},
       eprint = {2306.16828},
 primaryClass = {astro-ph.CO},
       adsurl = {https://ui.adsabs.harvard.edu/abs/2023Ap&SS.368...59F}
}


\begin{appendix}
\section{Spectroscopic observations of discrepant redshifts}
\label{appendix:discrepancies-spectra}

12 QSOs with redshift discrepancies between the literature and the present Gaia estimate (Tab. \ref{tab:completeness_outliers_observed}) have been observed with follow-up spectroscopy at the Southern Astrophysical Research (SOAR) and ESO New Technology Telescope (ESO-NTT) telescopes using the Goodman High-Throughput Spectrograph \citep{GOODMAN2004} and ESO Faint Object Spectrograph 2 \citep[EFOSC2, ][]{EFOSC:1984Msngr..38....9B} instruments, respectively. 

Fig. \ref{Fig:outlier_spectra} shows the spectra obtained from the observations.

7 candidates were observed in September and October 2024 at the SOAR telescope, using the Goodman High Throughput Spectrograph with the Red camera and the $400 l/mm$ VPH Grating (wavelength range $\lambda\sim 5000-9000$ {\AA} and resolution $\sim 1850$), with exposure times between 600 and 900 s.
4 more candidates were observed with Goodman in May 2025, using the Blue camera and the $400 l/mm$ VPH Grating (wavelength range $\lambda\sim 3000-7050$ {\AA} and resolution $\sim 1850$). All Goodman data were obtained during engineering time.

1 candidate was observed in November 2024, as a part of an observing program at the ESO NTT (PI. A. Grazian, proposal 114.27HT.001), employing the EFOSC-2 instrument and Grism \#13 (wavelength range $\lambda\sim 3700-9300$ {\AA} and resolution $\sim 1000$), with exposure time of 600 s.

Data obtained with EFOSC-2 were reduced with a custom pipeline based on MIDAS scripts \citep{MIDAS1988}. Each spectrum has been processed to subtract the bias and normalised by the flat; wavelength calibration is achieved using helium, neon and argon lamps, finding a rms of $\sim0.5${\AA}; flux calibration was performed using spectroscopic flux standards observed at the beginning of the night.

Data obtained with Goodman HTS were reduced using the custom Goodman reduction pipeline \citep{GOODMAN_Pipeline2020} that applies bias subtraction, flat field correction and wavelength calibration to each individual science frame. Flux calibration was performed using spectroscopic flux standards observed at the beginning of the night. 
Observing conditions have not always been photometric.

All 12 QSOs turned out to have a spectroscopic redshift in excellent agreement with the redshift from the Gaia low-resolution spectra (Table \ref{tab:completeness_outliers_observed}), thus further confirming the validity of the independent sample.

\begin{table*}[bp]
\centering
\caption{Spectroscopic observations of 12 QSO with discrepant redshift.}
\begin{tabular}{|r|c|c|c|c|c|c|c|c|c|c|}
\hline
\textbf{qid} & \textbf{RA} & \textbf{DEC} & \textbf{Gmag} & \textbf{z\_spec} & \textbf{z\_QU\_G} & \textbf{z\_spec\_obs} & \textbf{obs\_date} & \textbf{exptime} & \textbf{instrument} \\ \hline
873205  & 22:01:21.23 & -32:31:44.6 & 17.75 & 2.13  & 1.996 & 2.009 & 2024-09-14 & 600 s & Goodman R \\ \hline
896171  & 21:57:54.06 & -34:21:00.3 & 17.17 & 1.88  & 2.020 & 2.017 & 2024-09-14 & 600 s & Goodman R \\ \hline
917239  & 22:36:33.62 & -46:44:00.0 & 17.29 & 2.3   & 2.143 & 2.145 & 2024-09-14 & 600 s & Goodman R \\ \hline
2226338 & 21:50:15.34 & -37:01:42.1 & 17.86 & 2.7   & 2.608 & 2.610 & 2024-09-14 & 900 s & Goodman R \\ \hline
884515  & 21:25:05.85 & -69:57:56.5 & 17.93 & 2.103 & 2.011 & 2.015 & 2024-10-21 & 900 s & Goodman R \\ \hline
1080913 & 03:43:59.36 & -38:33:05.8 & 17.89 & 2.45  & 2.319 & 2.322 & 2024-10-21 & 900 s & Goodman R \\ \hline
811314  & 03:32:44.11 & -44:55:57.3 & 17.46 & 2.6   & 2.676 & 2.670 & 2024-10-21 & 900 s & Goodman R \\ \hline
1155758 & 22:13:45.11 & -34:00:51.9 & 18.06 & 2.4   & 2.320 & 2.325 & 2024-11-17 & 600 s & EFOSC2    \\ \hline
1053453 & 14:16:42.31 & -21:31:55.0 & 17.71 & 2.3   & 2.180 & 2.183 & 2025-05-03 & 900 s & Goodman B \\ \hline
1141942 & 11:31:18.62 & -29:07:15.9 & 18.24 & 2.3   & 2.192 & 2.191 & 2025-05-03 & 900 s & Goodman B \\ \hline
1141944 & 11:31:51.10 & -29:18:14.8 & 18.21 & 2.65  & 2.733 & 2.688 & 2025-05-03 & 900 s & Goodman B \\ \hline
2220198 & 13:17:44.20 & -31:47:13.0 & 18.16 & 3.1   & 2.951 & 2.940 & 2025-05-03 & 900 s & Goodman B \\ \hline
\end{tabular}
\tablefoot{Observed objects with redshift discrepancies between the literature ($z_{\rm spec}$) and the present Gaia estimate ($z_{\rm QU\_G}$). Observations ($z_{\rm spec\_obs}$) confirm in all cases the Gaia estimates.}
\label{tab:completeness_outliers_observed}
\end{table*}

\begin{figure*}[hp]
    \centering
    \includegraphics[width=0.84\textwidth]{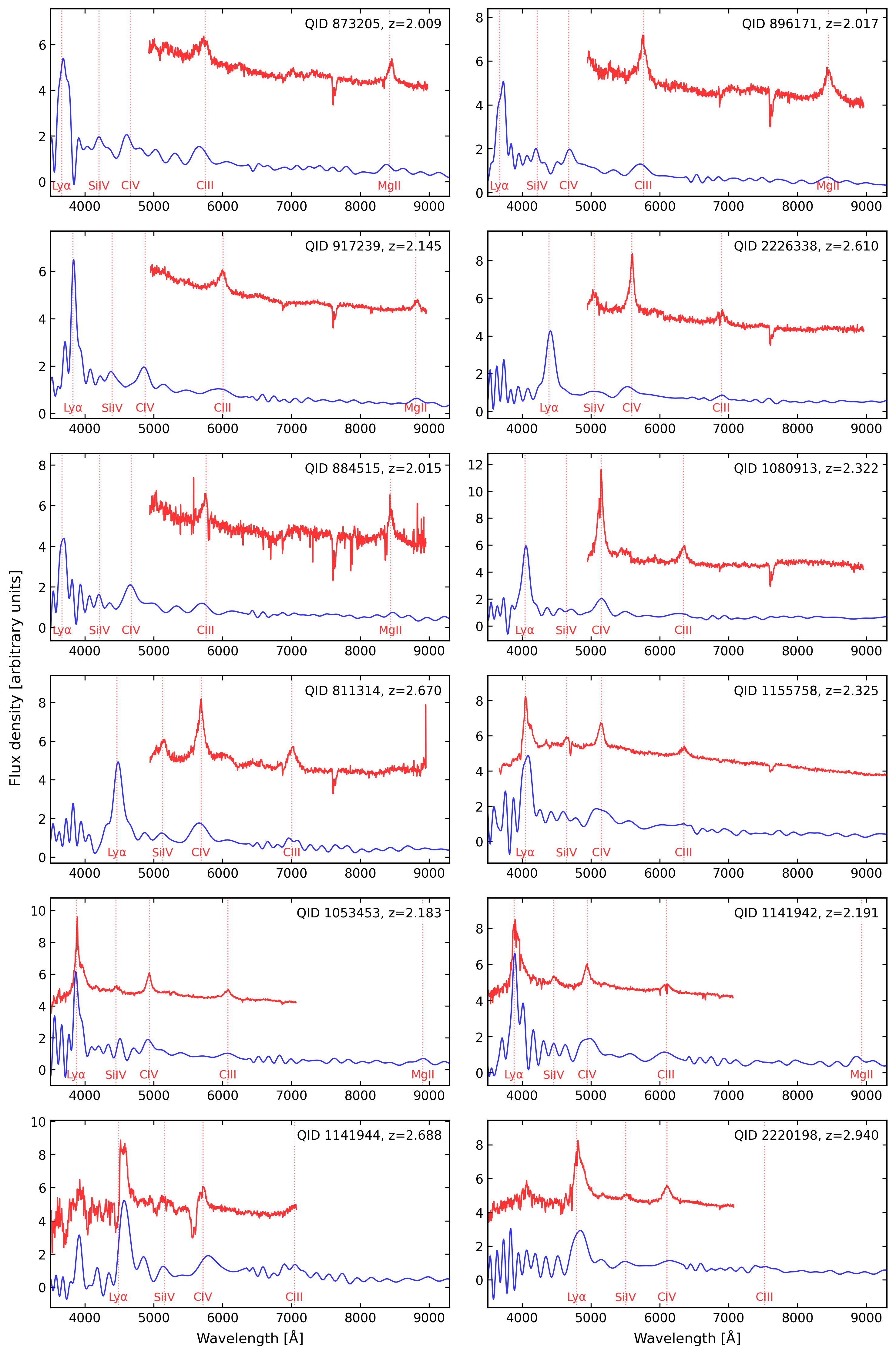}
    \caption{Gaia spectra (in blue) and Goodman/EFOSC2 spectra (in red) for the 12 observed objects with redshift discrepancies between the literature and the present Gaia estimate (see Table \ref{tab:completeness_outliers_observed}). Observations confirm in all cases the Gaia estimates.}
    \label{Fig:outlier_spectra}
\end{figure*}
\newpage
\section{Examples of Gaia DR3 Spectra}
\label{appendix:gaia-spectra-examples}

Fig. \ref{Fig:gaia-spectra-qop-comparison} contains examples of 12 Gaia DR3 spectra showing the four QOP quality levels used in this work (see Section \ref{sec:define_qso_sample}). QOP=1 (top): uncertain identification with low SNR and ambiguous features; QOP=2 (second): acceptable quality with identifiable but weak emission lines; QOP=3 (third): good quality with clear emission lines enabling secure classification; QOP=4 (bottom): excellent quality with strong emission lines and high SNR. Only QOP$\geq$2 objects were included in our analysis to ensure reliable redshift measurements.

\begin{figure*}[hp]
    \centering
    \includegraphics[width=0.84\textwidth]{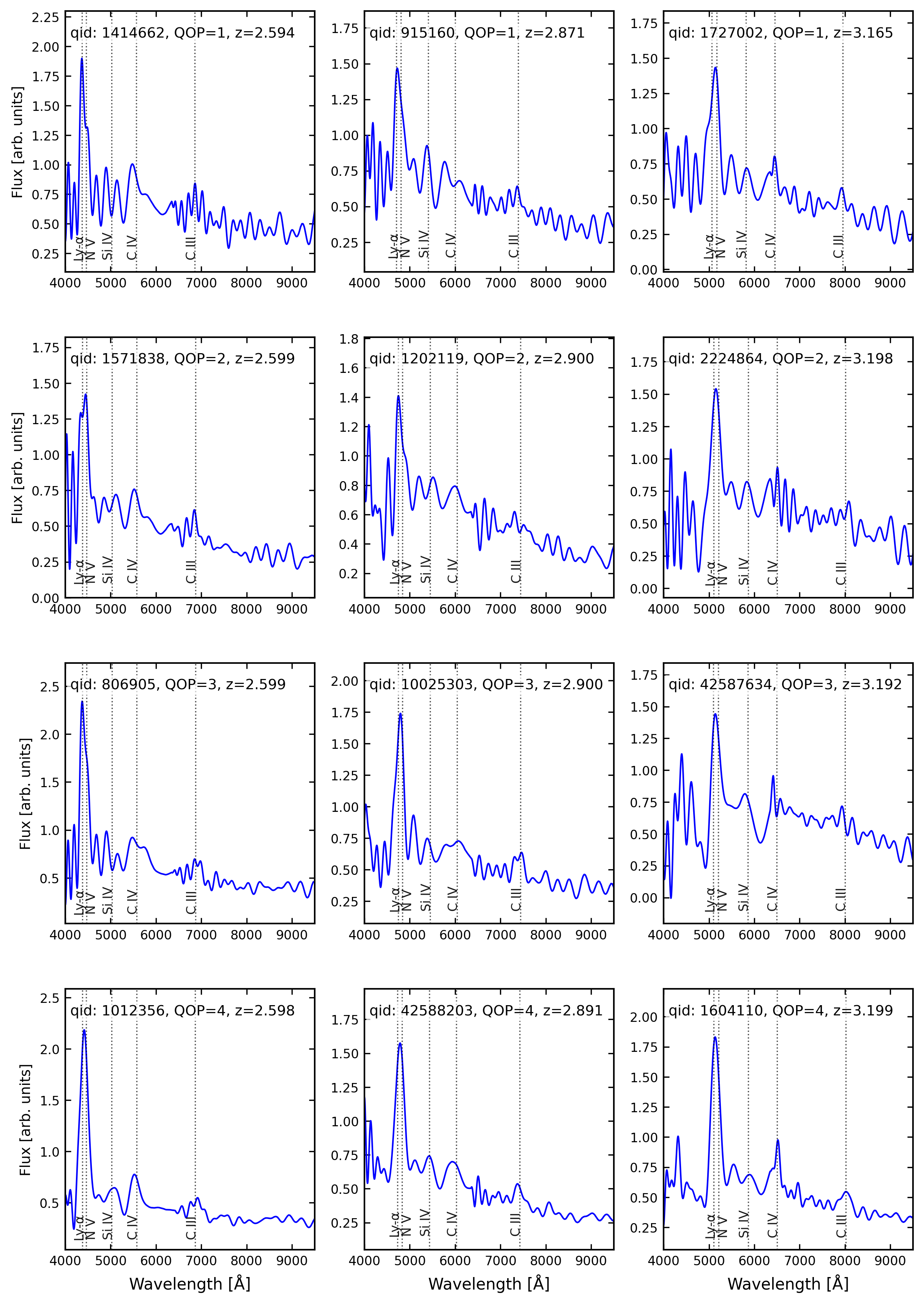}
    \caption{Gaia DR3 spectra of different quality: QOP=1 for the first row, QOP=2 for the second, QOP=3 for the third and QOP=4 for the fourth.}
    \label{Fig:gaia-spectra-qop-comparison}
\end{figure*}

A further proof of the quality of Gaia spectra is shown in Fig. \ref{Fig:qso_sample_spectra}, that contains the Gaia DR3 spectra of the 12 QSOs with the highest redshift among the new identifications obtained in this work (corresponding to the objects listed in Table \ref{tab:qso_sample}).

\begin{figure*}[hp!]
    \centering
    \includegraphics[width=0.8\textwidth]{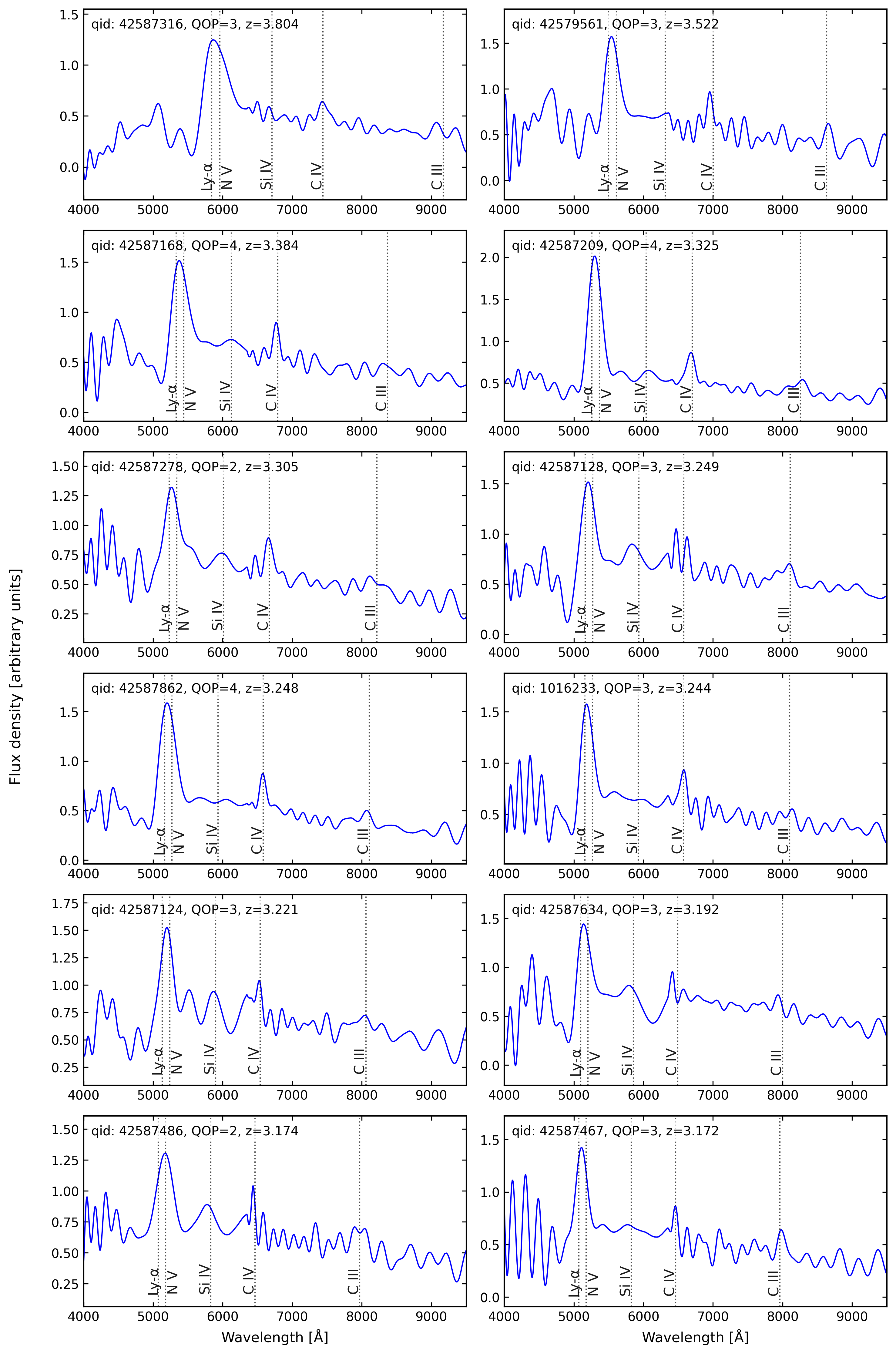}
    \caption{Gaia spectra of the 12 highest-redshift QSOs discovered in this work (see Table \ref{tab:qso_sample}).}
    \label{Fig:qso_sample_spectra}
\end{figure*}

\clearpage

\section{Data availability}
Table \ref{tab:qso_sample} is only available in electronic form at the CDS via anonymous ftp to cdsarc.u-strasbg.fr (130.79.128.5) or via \url{http://cdsweb.u-strasbg.fr/cgi-bin/qcat?J/A+A/}.

\section{Acknowledgements}
\begin{acknowledgements}
Andrea Grazian, Stefano Cristiani, Andrea Trost, Valentina D'Odorico, Giorgio Calderone, and Matteo Porru acknowledge the financial support of the INAF GO/GTO Grant 2023 ``Finding the Brightest Cosmic Beacons in the Universe with QUBRICS'' (PI Grazian), and of the Italian Ministry of Education, University, and Research with PRIN 201278X4FL and the "Progetti Premiali" funding scheme. 
Stefano Cristiani is partly supported by the INFN PD51 INDARK grant. 
The work of Konstantina Boutsia is supported by NOIRLab, which is managed by the Association of Universities for Research in Astronomy (AURA) under a cooperative agreement with the U.S. National Science Foundation.
Catarina Marques acknowledges the supported by FCT - Fundação para a Ciência e Tecnologia, I.P. by project reference 2023.03984.BD and DOI identifier https://doi.org/10.54499/2023.03984.BD.
This work has made use of observations collected at the European Southern Observatory under ESO programme(s) 114.27HT.001. 
This research has made use of the SIMBAD database, CDS, Strasbourg Astronomical Observatory, France \citep{SIMBAD:2000A&AS..143....9W}.
This research has made use of the NASA/IPAC Extragalactic Database, which is funded by the National Aeronautics and Space Administration and operated by the California Institute of Technology.
This work has made use of data from the European Space Agency (ESA) mission {\it Gaia}(\url{https://www.cosmos.esa.int/gaia}), processed by the {\it Gaia} Data Processing and Analysis Consortium (DPAC, \url{https://www.cosmos.esa.int/web/gaia/dpac/consortium}). Funding for the DPAC has been provided by national institutions, in particular the institutions participating in the {\it Gaia} Multilateral Agreement.
This publication makes use of data products from the Wide-field Infrared Survey Explorer, which is a joint project of the University of California, Los Angeles, and the Jet Propulsion Laboratory/California Institute of Technology, funded by the National Aeronautics and Space Administration.
This publication makes use of data products from the Near-Earth Object Wide-field Infrared Survey Explorer (NEOWISE), which is a joint project of the Jet Propulsion Laboratory/California Institute of Technology and the University of California, Los Angeles. NEOWISE is funded by the National Aeronautics and Space Administration.
The Pan-STARRS1 Surveys (PS1) and the PS1 public science archive have been made possible through contributions by the Institute for Astronomy, the University of Hawaii, the Pan-STARRS Project Office, the Max-Planck Society and its participating institutes, the Max Planck Institute for Astronomy, Heidelberg and the Max Planck Institute for Extraterrestrial Physics, Garching, The Johns Hopkins University, Durham University, the University of Edinburgh, the Queen's University Belfast, the Harvard-Smithsonian Center for Astrophysics, the Las Cumbres Observatory Global Telescope Network Incorporated, the National Central University of Taiwan, the Space Telescope Science Institute, the National Aeronautics and Space Administration under Grant No. NNX08AR22G issued through the Planetary Science Division of the NASA Science Mission Directorate, the National Science Foundation Grant No. AST–1238877, the University of Maryland, Eotvos Lorand University (ELTE), the Los Alamos National Laboratory, and the Gordon and Betty Moore Foundation.
The national facility capability for SkyMapper has been funded through ARC LIEF grant LE130100104 from the Australian Research Council, awarded to the University of Sydney, the Australian National University, Swinburne University of Technology, the University of Queensland, the University of Western Australia, the University of Melbourne, Curtin University of Technology, Monash University and the Australian Astronomical Observatory. SkyMapper is owned and operated by The Australian National University's Research School of Astronomy and Astrophysics. The survey data were processed and provided by the SkyMapper Team at ANU. The SkyMapper node of the All-Sky Virtual Observatory (ASVO) is hosted at the National Computational Infrastructure (NCI). Development and support of the SkyMapper node of the ASVO has been funded in part by Astronomy Australia Limited (AAL) and the Australian Government through the Commonwealth's Education Investment Fund (EIF) and National Collaborative Research Infrastructure Strategy (NCRIS), particularly the National eResearch Collaboration Tools and Resources (NeCTAR) and the Australian National Data Service Projects (ANDS).
This project used data obtained with the Dark Energy Camera (DECam), which was constructed by the Dark Energy Survey (DES) collaboration. Funding for the DES Projects has been provided by the U.S. Department of Energy, the U.S. National Science Foundation, the Ministry of Science and Education of Spain, the Science and Technology Facilities Council of the United Kingdom, the Higher Education Funding Council for England, the National Center for Supercomputing Applications at the University of Illinois at Urbana-Champaign, the Kavli Institute of Cosmological Physics at the University of Chicago, Center for Cosmology and Astro-Particle Physics at the Ohio State University, the Mitchell Institute for Fundamental Physics and Astronomy at Texas A\&M University, Financiadora de Estudos e Projetos, Fundacao Carlos Chagas Filho de Amparo, Financiadora de Estudos e Projetos, Fundacao Carlos Chagas Filho de Amparo a Pesquisa do Estado do Rio de Janeiro, Conselho Nacional de Desenvolvimento Cientifico e Tecnologico and the Ministerio da Ciencia, Tecnologia e Inovacao, the Deutsche Forschungsgemeinschaft and the Collaborating Institutions in the Dark Energy Survey. The Collaborating Institutions are Argonne National Laboratory, the University of California at Santa Cruz, the University of Cambridge, Centro de Investigaciones Energeticas, Medioambientales y Tecnologicas-Madrid, the University of Chicago, University College London, the DES-Brazil Consortium, the University of Edinburgh, the Eidgenossische Technische Hochschule (ETH) Zurich, Fermi National Accelerator Laboratory, the University of Illinois at Urbana-Champaign, the Institut de Ciencies de l’Espai (IEEC/CSIC), the Institut de Fisica d’Altes Energies, Lawrence Berkeley National Laboratory, the Ludwig Maximilians Universitat Munchen and the associated Excellence Cluster Universe, the University of Michigan, NSF’s NOIRLab, the University of Nottingham, the Ohio State University, the University of Pennsylvania, the University of Portsmouth, SLAC National Accelerator Laboratory, Stanford University, the University of Sussex, and Texas A\&M University.
This research used data obtained with the Dark Energy Spectroscopic Instrument (DESI). DESI construction and operations is managed by the Lawrence Berkeley National Laboratory. This material is based upon work supported by the U.S. Department of Energy, Office of Science, Office of High-Energy Physics, under Contract No. DE–AC02–05CH11231, and by the National Energy Research Scientific Computing Center, a DOE Office of Science User Facility under the same contract. Additional support for DESI was provided by the U.S. National Science Foundation (NSF), Division of Astronomical Sciences under Contract No. AST-0950945 to the NSF’s National Optical-Infrared Astronomy Research Laboratory; the Science and Technology Facilities Council of the United Kingdom; the Gordon and Betty Moore Foundation; the Heising-Simons Foundation; the French Alternative Energies and Atomic Energy Commission (CEA); the National Council of Humanities, Science and Technology of Mexico (CONAHCYT); the Ministry of Science and Innovation of Spain (MICINN), and by the DESI Member Institutions: www.desi.lbl.gov/collaborating-institutions. The DESI collaboration is honored to be permitted to conduct scientific research on I’oligam Du’ag (Kitt Peak), a mountain with particular significance to the Tohono O’odham Nation. Any opinions, findings, and conclusions or recommendations expressed in this material are those of the author(s) and do not necessarily reflect the views of the U.S. National Science Foundation, the U.S. Department of Energy, or any of the listed funding agencies.

\end{acknowledgements}

\end{appendix}

\end{document}